\documentclass[12pt]{article}
\pdfoutput=1
\usepackage{jheppub}

\usepackage{amssymb,amsmath}
\usepackage{graphicx}
\usepackage{subcaption} 
\usepackage{bm}
\usepackage{epsfig}
\usepackage{epstopdf}
\usepackage{latexsym}
\usepackage{psfrag}
\usepackage{booktabs}
\usepackage{bbm}

\usepackage{color}
\usepackage{datetime}

\newcommand{\pref}[1]{(\ref{#1})}

\def\({\left(}
\def\){\right)}
\def\[{\left[}
\def\]{\right]}

\def\eg{{e.g.}}

\def\etc{{etc}}

\def\lstr{\ell_{\rm str}}
\def\gstr{g_{\rm s}}

\def\half{\frac12}
\def\coeff#1#2{{\textstyle \frac{#1}{#2}}}

\def\One{{\hbox{ 1\kern-.8mm l}}}

\def\barray{\begin{array}}
\def\earray{\end{array}}
\def\be{\begin{equation}}
\def\ee{\end{equation}}
\def\bea{\begin{eqnarray}}
\def\eea{\end{eqnarray}}
\def\bal{\begin{align}}
\def\eal{\end{align}}
\def\nn{\nonumber}

\def\xx{{\bf x}}
\def\yy{{\bf y}}
\def\yscale{{{\bf \hat y}}}
\def\zscale{{\hat z}}

\def\Da{{\mathcal A}}

\def\psib{{\boldsymbol\psi}}
\def\vb{{\boldsymbol v}}

\def\coeff#1#2{\relax{\textstyle {\frac {#1}{#2}}}\displaystyle}

\def\RR{\mathbb{R}}
\def\SS{\mathbb{S}}
\def\TT{\mathbb{T}}
\def\ZZ{\mathbb{Z}}
\def\cA{{\cal A}}
\def\cB{{\cal B}}
\def\cC{{\cal C}}

\def\cF{{\cal F}}
\def\cG{{\cal G}}
\def\cI{{\cal I}}

\def\cM{{\cal M}}

\def\cO{{\cal O}}

\def\cQ{{\cal Q}}
\def\cR{{\cal R}}
\def\cS{{\cal S}}

\def\cW{{\cal W}}

\definecolor{cardinal}{rgb}{0.6,0,0}
\definecolor{darkgreen}{rgb}{0,0.5,0}
\definecolor{golden}{rgb}{0.92, 0.7, 0}
\definecolor{midnight}{rgb}{0, 0, 0.5}
\definecolor{darkblue}{rgb}{0.2, 0, 0.8}
\definecolor{white}{rgb}{1,1,1}
\definecolor{black}{rgb}{0,0,0}

\newcommand{\dd}{\mathrm{d}}					
\DeclareMathOperator*{\hodge}{\star}				
\DeclareMathOperator*{\thodge}{\tilde{\star}}		
\newcommand{\vol}{\mathrm{vol}}				

\newcommand{\norm}[1]{{\lVert {#1} \rVert}}			
\newcommand{\bnorm}[1]{{\big \lVert {#1} \big \rVert}}
\newcommand{\abs}[1]{{\lvert {#1} \rvert}}			

\usepackage{forloop}
\newcounter{ct}

\newcommand{\eucl}[1]{(%
	\ifthenelse{#1 > 0}{%
		\mathord{+} \forloop[-1]{ct}{#1}{\value{ct} > 1}{\, \mathord{+}}%
	}{}%
)}

\newcommand{\sig}[2]{(%
	\ifthenelse{#2 > 0}{%
		\mathord{-}  \forloop[-1]{ct}{#2}{\value{ct} > 1}{\, \mathord{-}}%
		\ifthenelse{#1 > 0}{\,}{}%
	}{}%
	\ifthenelse{#1 > 0}{%
		\mathord{+} \forloop[-1]{ct}{#1}{\value{ct} > 1}{\, \mathord{+}}%
	}{}%
)}

\usepackage{empheq}
\newcommand*\padbox[1]{\fbox{\hspace{0.3em}#1\hspace{0.3em}}}
\empheqset{box=\padbox}

\newcommand{\frv}{\vec{\mathfrak{u}}}		
\newcommand{\fra}{\vec{\mathfrak{a}}}		

\allowdisplaybreaks[1] 

\begin{document}  

 
\title{
Hair-brane Ideas on the Horizon
}

\author[a]{Emil J. Martinec}
\author[b]{{\it and\;} Ben E. Niehoff}

\affiliation[a]{Enrico Fermi Institute and Department of Physics, University of Chicago\\ 5640 S. Ellis Ave., Chicago, IL 60637-1433, USA}
\affiliation[b]{Dept. of Applied Mathematics and Theoretical Physics, University of Cambridge \\ Centre for Mathematical Sciences, Wilberforce Rd., Cambridge, CB3 0WA, United Kingdom}

\emailAdd{e-martinec@uchicago.edu} 
\emailAdd{B.E.Niehoff@damtp.cam.ac.uk}



\abstract{
We continue an examination of the microstate geometries program begun in 
arXiv:1409.6017, focussing on the role of branes that wrap the cycles which degenerate when a throat in the geometry deepens and a horizon forms.  An associated quiver quantum mechanical model of minimally wrapped branes exhibits a non-negligible fraction of the gravitational entropy, which scales correctly as a function of the charges.  The results suggest a picture of $\rm AdS_3/CFT_2$ duality wherein the long string that accounts for BTZ black hole entropy in the CFT description, can also be seen to inhabit the horizon of BPS black holes on the gravity side.
}

\maketitle



\section{Introduction}

Gauge/gravity duality is a weak/strong coupling duality.
For instance, quantum gravity on $AdS_3\times \SS^3\times \TT^4$ is dual to the 1+1d symmetric orbifold CFT $(\TT^4)^N\!/S^N$; the symmetric orbifold lives in a cusp of the moduli space where the ambient curvature of the geometry is string scale%
~\cite{Larsen:1999uk},
while weakly curved $AdS$ spacetime lies in the middle of the moduli space.  
The invariance of the BPS spectrum under deformations of the moduli guarantees that one can enumerate the microstates that account for the gravitational entropy of BPS black holes via a weak-coupling calculation%
~\cite{Strominger:1996sh}.   
The asymptotic Virasoro symmetry of $AdS_3$ spacetime further guarantees that, within a framework of unitary dynamics, one has an accounting of the asymptotic non-BPS entropy%
~\cite{Strominger:1997eq}, 
and thus in principle a resolution of the information paradox for black holes%
~\cite{Hawking:1976ra}.

However, using the duality to answer specific questions about local geometry of black holes, and thereby gain an understanding of the {\it mechanism} by which the information paradox is resolved, is a daunting task for which the CFT is ill-suited.  First one needs to reconstruct all of spacetime in the gauge theory, then insert a black hole, then probe its geometry near/beyond the horizon.

It has been suggested that, since a microstate itself has no degeneracy, one might try to approach the problem from the other direction~-- to match the CFT accounting by enumerating horizonless solutions to the supergravity field equations with the same asymptotic charges as the black hole states (for an early review, see%
~\cite{Mathur:2005zp}).  
The idea here is that an individual microstate does not represent an ensemble of states and therefore should have no horizon.  Of course, it is not guaranteed that this program will succeed in this simplest realization, as the necessary states may lie beyond the supergravity regime~-- requiring for instance brane sources or involving string- or Planck-scale curvatures (for an overview of the strategy, see%
~\cite{Bena:2013dka}).

This {\it microstate geometries} program has had a number of successes, beginning with the complete enumeration of two-charge solutions to the field equations%
~\cite{Lunin:2001fv}. 
This led to the hope that a representative sample of microstates of macroscopic three-charge black holes could be obtained by enumerating horizonless three-charge solutions.  The past few years have seen remarkable progress in this program (for recent results and references to earlier work, see%
~\cite{%
Bena:2014qxa,%
Bena:2015bea,%
Giusto:2015dfa%
}).
Although current methods still fall short of exhibiting an entropy that scales with the same power of the charges as the BPS entropy
\be
\label{BPS entropy}
S = 2\pi \sqrt{ Q_1 Q_2 Q_3} ~,
\ee
increasingly sophisticated solution-generating techniques continue to be elaborated.

But are microstate geometries black holes?  In a recent work by one of the authors%
~\cite{Martinec:2014gka}, 
it was pointed out that a key ingredient of the construction of microstate geometries -- the slight separation of the background charge sources into a ``supertube'' configuration%
~\cite{Bena:2011uw} --
bears a strong analogy to the desingularization of the throat geometry of coincident NS5-branes by separating them slightly onto the Coulomb branch of their moduli space%
~\cite{Giveon:1999zm,Giveon:1999px,Giveon:1999tq}.  For $Q_1 = n_5$ NS5-branes compactified on a circle, the BPS entropy~\pref{BPS entropy} is accounted for by the BPS states of a {\it little string}, whose tension is $n_5$ times smaller than the fundamental string tension%
~\cite{Maldacena:1996ya},
and which carries winding $w=Q_2$ and momentum $p=Q_3$ around the circle.  The little string's tension becomes large sufficiently far out on the Coulomb branch, and so the light BPS spectrum there is accounted for instead by the elliptic genus of perturbative fundamental strings%
~\cite{Bena:2008dw,Bena:2010gg,Giveon:2015raa} 
in what they see as a smooth geometry.  This entropy is much less than that of equation~\pref{BPS entropy} when $n_5$ is large.

The entropy of three-charge black holes in asymptotically $AdS_3$ spacetime is similarly counted by the states of a {\it long string}%
~\cite{Strominger:1996sh}.  The suggestion of%
~\cite{Martinec:2014gka}
is that the nonsingular supertube configurations of the microstate geometries should be thought of as Coulomb branch states in which the long string is lifted out of the light spectrum; and that as in little string theory, one should look for the long string to appear in regions of the moduli space where sources coalesce.  From this perspective, microstate geometries join the wide variety of situations where singularities are resolved in string theory via the appearance of new, light degrees of freedom associated to extended objects (the prototypes being%
~\cite{Strominger:1995cz,Hull:1995mz}).  
The novel feature here is that a macroscopic number of states become light at the same time in order to account for the BPS black hole entropy.

The present work elaborates on this proposal.  
In section~\ref{sec:W-branes}, we review the structure of NS5-branes on the Coulomb branch, and the similar structures that appear in two-charge microstate geometries. 
In section~\ref{sec:SugraFrames}, we generalize the discussion to three-charge geometries.
The objects that become light when charge centers coalesce in microstate geometries are branes that wrap degenerating cycles threaded by form field magnetic flux.  In the spirit of%
~\cite{Strominger:1995cz,Hull:1995mz}, in section~\ref{sec:QuiverQM} we second-quantize the dynamics of branes that minimally wrap these cycles using the quiver quantum mechanical model of%
~\cite{Denef:2002ru}. 
The resulting BPS entropy of the quiver%
~\cite{Denef:2007vg,Bena:2012hf}
represents a nontrivial fraction of the BPS states of three-charge black holes/black rings and scales in the same way as a function of the charges.
Section~\ref{sec:probes} derives the data of the quiver from the dynamics of probe branes wrapping the relevant cycles from the perspective of both M-theory and type IIB duality frames.  We conclude with a discussion in section~\ref{sec:Discussion}.  
Three appendices contain details of the calculations: Appendix~\ref{T duality conventions} summarizes our conventions for duality transformations and probe brane effective actions; Appendix~\ref{6d geom details} elaborates some details of the IIB geometry, as well as the simplest example, namely the geometries with two charge centers; and Appendix~\ref{M2-D3 duality} shows how the D3-brane gauge field dynamics in the type IIB background is related to the motion of the M2-brane in the dual M-theory background.


\section{Singularities, scaling limits, and W-branes}\label{sec:W-branes}

\begin{figure}
\centerline{\includegraphics[width=3.5in]{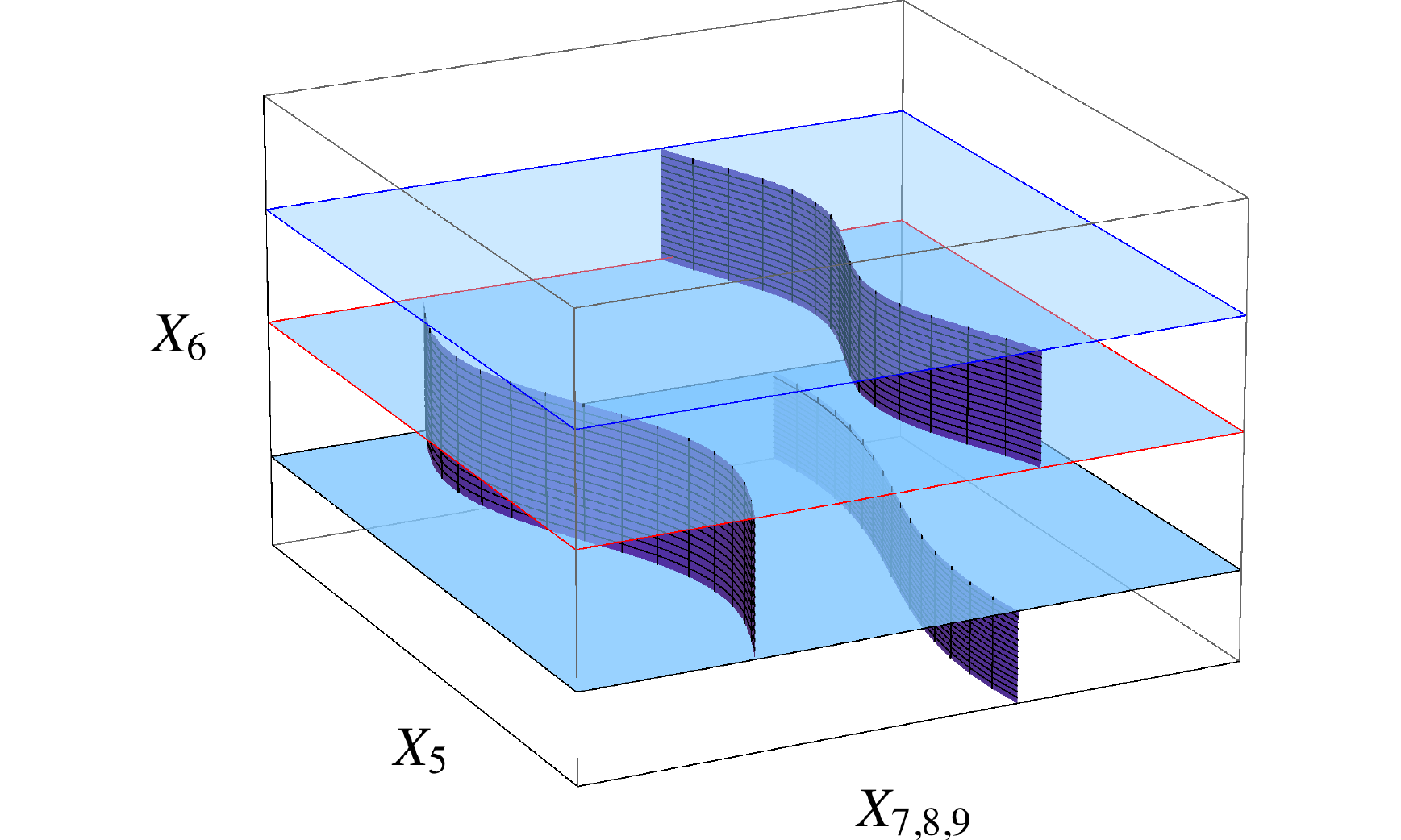}}
\setlength{\unitlength}{0.1\columnwidth}
\caption{\it 
A D2-brane fractionates on a stack of D4-branes.  U-duality and scaling limits relate this situation to both $AdS_3\times\SS^3\times\TT^4$ and little string theory.}
\label{fig:wiggles}
\end{figure}

Prior to taking the scaling limit that leads to an asymptotically $AdS_3\times \SS^3\times \TT^4$ spacetime, three-charge BPS geometries are continuously related via a different scaling limit to the BPS spectrum of little string theory (for a review of little string theory, see%
~\cite{Kutasov:2001uf}).  
Consider the D2-D4 system in type IIA which is T-dual to D1-D5 on $\TT^5$.  At weak coupling, a D2-brane bound to a stack of D4-branes fractionates into $n_4$ strips whose boundary intersection curves along the D4-branes are effective string-like degrees of freedom, see figure~\ref{fig:wiggles}.  
If the D2-branes wrap a circle transverse to the D4's, we can exchange this circle with the M-theory circle in which case the D4's become NS5-branes and the fractional D2-branes become fractional fundamental strings.  Thus the BPS spectrum of the D2-D4 system matches that of little string theory compactified on $\TT^5$%
~\cite{Dijkgraaf:1996cv}.  
However, the nonextremal limit differs, because the torus is scaled differently in the decoupling limit~-- for little string theory, one holds all the torus cycles fixed in string units, and then takes the asymptotic string coupling $\gstr\to 0$ in the duality frame where the background charges are NS5-F1.  For the AdS3 decoupling limit, one wraps the D2-D4 intersection around a torus cycle, parametrized (say) by a coordinate $v$ and with radius $R_v$; then the transverse $\TT^4$ is held fixed in string units while taking the IR limit $\lstr E\to 0$, holding $E R_v$ fixed.

\subsection{The worldsheet description of little string theory} 

Given this close connection between little string theory and $AdS_3$ backgrounds, one might look to the former for intuition about the latter.  The decoupling limit that leads to little string theory was in fact an early instance of a near-horizon limit%
~\cite{Callan:1991dj}.  
Strings propagating in the (CHS) throat of $n_5$ coincident NS5-branes is described by a free linear dilaton CFT for the radial direction of the throat, together with an $SU(2)$ level $n_5$  WZNW theory for the angular directions (and whatever CFT describes the directions tangent to the branes).  The worldsheet theory is singular, as strings propagate freely to the large coupling down the throat where perturbation theory breaks down.

Separating the NS5-branes on the Coulomb branch moduli space renders the worldsheet perturbation theory finite and well-behaved%
~\cite{Giveon:1999zm,Giveon:1999px,Giveon:1999tq}.
For example, placing the branes along a contractible circle in $\ZZ_{n5}$ symmetric fashion replaces the CHS throat CFT by a nonsingular $\bigl(\frac{SL(2,\RR)}{U(1)}\times \frac{SU(2)}{U(1)}\bigr)/\ZZ_{n5}$ CFT.  The technical reason that string propagation is self-consistently weakly coupled when fivebranes are sufficiently separated is that fundamental strings cannot fit down the throat of a single fivebrane -- the minimal current algebra level of the supersymmetric $SU(2)$ WZNW model is two, due to the contribution of the worldsheet fermions.  Thus, while two or more coincident NS5-branes form a throat with a region of strong coupling that can be probed by fundamental strings, isolated fivebranes have no such throat,%
\footnote{Or at least, the description of the vicinity of an individual fivebrane is highly stringy, and not well described by a throat geometry.}
and the effective string coupling can only grow to a size governed by the distance to the nearest other fivebrane.

Consequently, the effective geometry in the decoupling limit of separated fivebranes is that of a capped throat.  In the case of fivebranes in a circular array, the radius of the circle sets a scale $\mu$ which determines the depth of the capped throat. 
More generally, the background of separated NS5-branes in supergravity is specified by a harmonic function $H$
\begin{align}
\dd s^2 &= -\dd t^2 + H(\xx)\,\dd \xx\cdot \dd \xx + \dd s^2_{{\mathbb T}^5}
\nonumber\\
H_{ijk} &= {\epsilon_{ijk}}^l \partial_l\, \log\bigl(H(\xx)\bigr)
\nonumber\\
e^{2\Phi} &= \gstr^2 H(\xx)
\\
H(\xx) &= \sum_{a=1}^{n_5} \frac{\alpha'}{|\xx-\xx_a|^2}
\nonumber ~,
\end{align}
and the separations $|\xx_b-\xx_a|$ govern the effective depth of the throat and therefore the weakness of the effective string coupling.  The apparent divergence of the string coupling as $\xx\to \xx_a$ is a red herring, as low-energy strings cannot visit this region.

D-branes can, however, resolve the throats of individual NS5-branes; D-branes can stretch between any pair $(a,b)$ of NS5-branes (see figure~\ref{fig:wiggles}, where now the vertical direction is the transverse space $x_{1,2,3,4}$, and the horizontal plane spanned by $x_5$ and $x_{6,7,8,9}$ is longitudinal to the fivebranes).  Perturbative string scattering processes are well-behaved at energy scales that don't excite the D-brane spectrum.

These Dirichlet ``W-branes'' are also related to the little string, directly in type IIA and indirectly in type IIB.  In type IIA, D2-branes stretch between NS5's, and both they and the little string source the tensor gauge theory on the fivebranes; the only difference is that the little string carries fractional F1 charge while the D2 W-branes do not (unless they carry fractional electric charge on their worldvolume).  
In type IIB, D1-branes stretch between NS5's and are the W-bosons of the fivebrane Yang-Mills gauge theory; the little string is a fractional instanton in this gauge theory, and thus a solitonic object.

At sufficiently small separation $|\xx_b-\xx_a|$, these stretched W-branes become lighter than fundamental strings and worldsheet perturbation theory breaks down.  Dynamics passes to a nonabelian Coulomb phase, whose excitations include the little string whose tension is $n_5$ times smaller than that of the fundamental string.

\subsection{Two-charge D1-D5 BPS microstate geometries}

Horizonless microstate geometries exhibit a structure qualitatively similar to little string theory on its Coulomb branch.  Consider for instance the two-charge BPS geometries, enumerated in%
~\cite{Lunin:2001fv,Lunin:2002bj}.
U-duality maps D1-D5 (or D2-D4) background charges to those of a fundamental string carrying winding charge $n_5$ and momentum $n_1$.  A BPS F1-P bound state is simply a string with oscillator excitations of only one chirality.  Anomaly cancellation requires that the total level of excitation is $N=n_1n_5$. 

The brane bound states in the U-duality orbit of the F1-P bound state are known as supertubes%
~\cite{Mateos:2001qs, Emparan:2001ux}.
They have the property that, in addition to conserved charges, the background supports dipole charge.  In the original F1-P duality frame, this simply results from the gyration of the string in the transverse plane due to the oscillator excitations; the bound state carries fundamental string dipole charge and angular momentum in the transverse plane. 

The geometry sourced by such a fundamental string was worked out explicitly in%
~\cite{Dabholkar:1995nc, Callan:1995hn}.
After mapping back to the D1-D5 duality frame, one sees that the dipole charges are  KK monopole charge and angular momentum, and it turns out that the geometry is completely smooth%
~\cite{Lunin:2002iz,Mathur:2005zp}, 
modulo orbifold singularities.  The geometries have a throat whose finite depth is governed by the typical separation of the strands of the gyrating KK monopole loop.  Their orientation varies along the loop, and thus attract one another; however the arrangement is held up by the centrifugal force of angular momentum.  Each oscillator contributes one-half to the angular momentum; the total angular momentum is a state in the tensor product  of doublets $2^{\otimes n}$, where $n$ is the number of oscillator quanta excited.

\begin{figure}[t]
\centering
\begin{subfigure}[b]{0.4\textwidth}
\centerline{\includegraphics[width=2.5in]{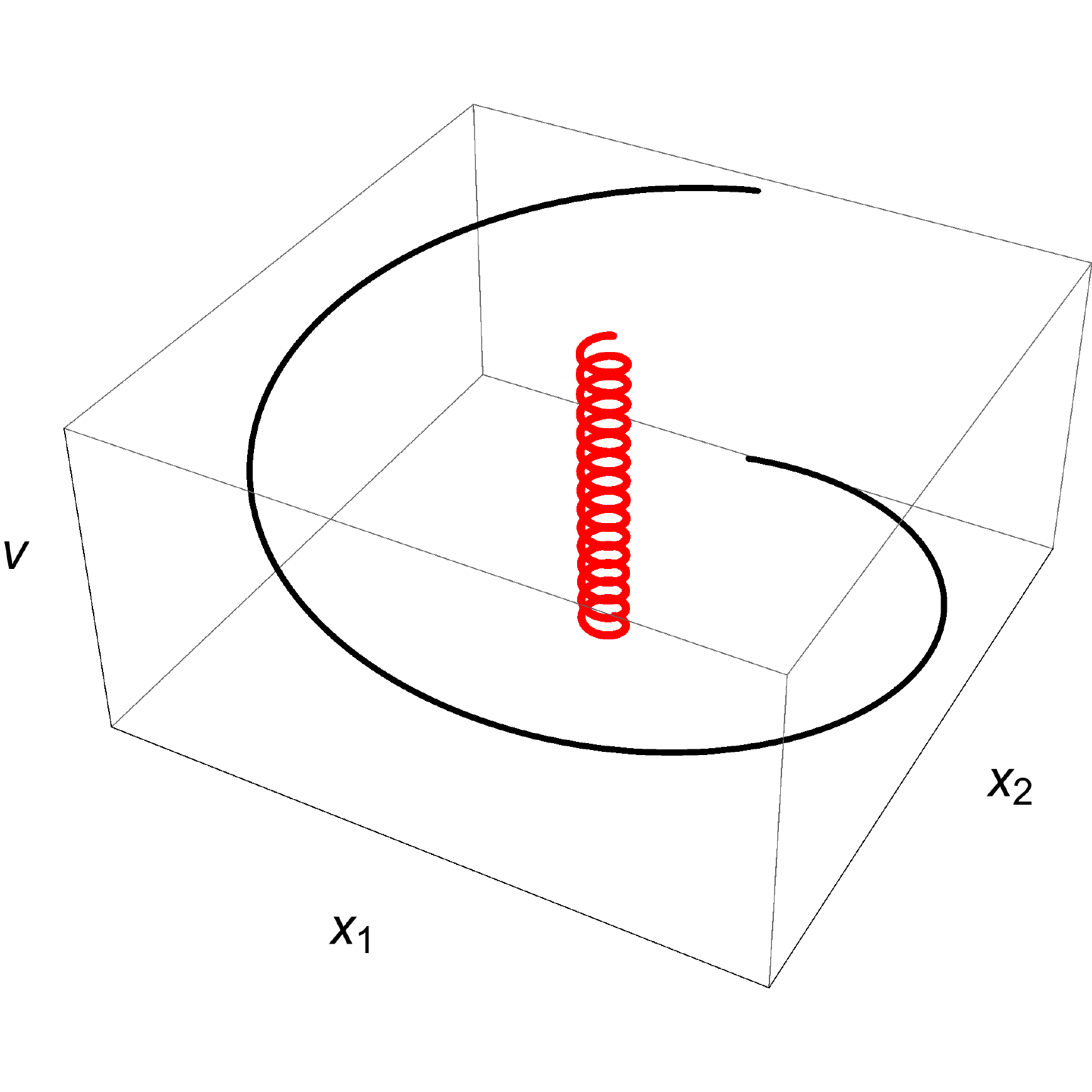}}
\end{subfigure}
\qquad
\begin{subfigure}[b]{0.4\textwidth}
\centerline{\includegraphics[width=2.5in]{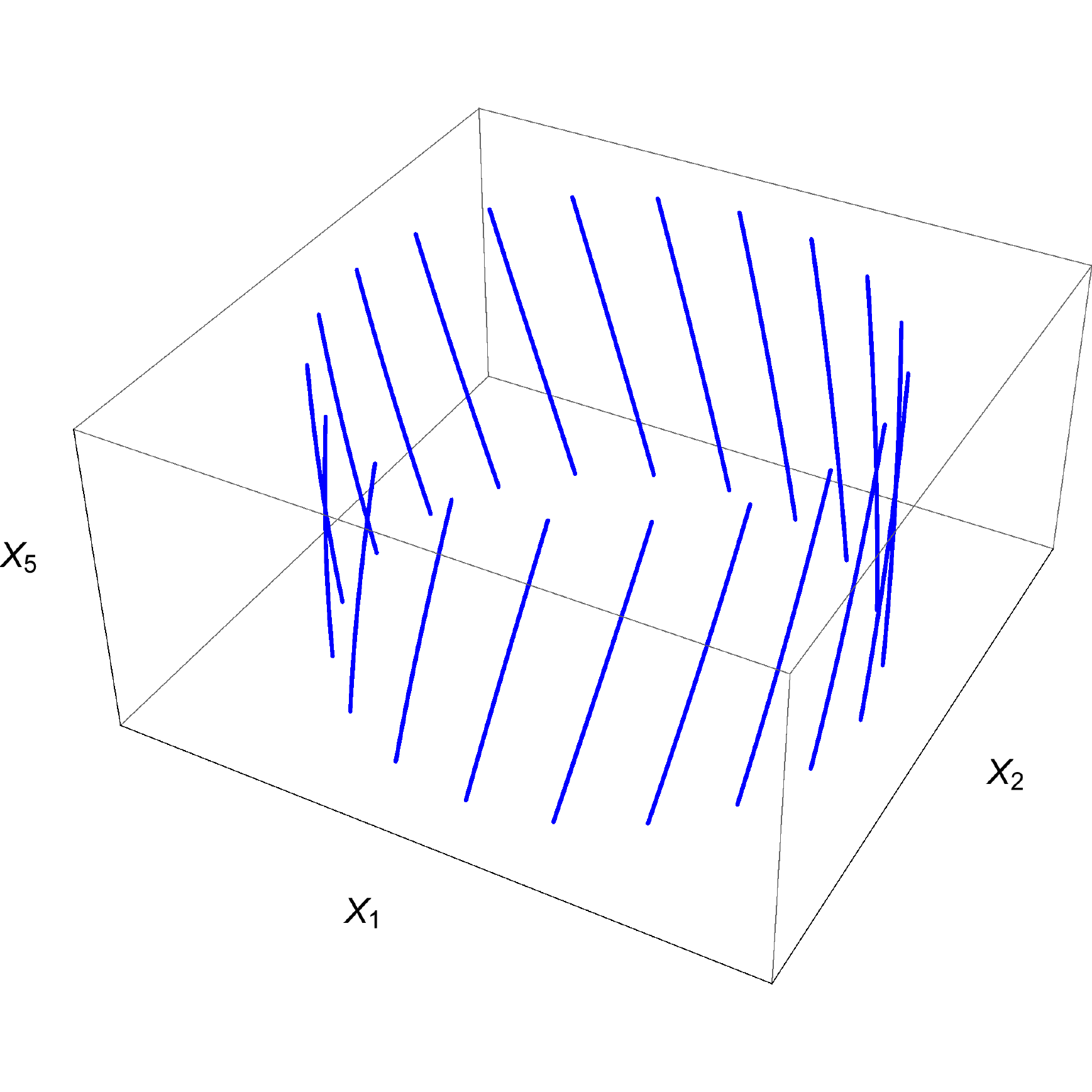}}
\end{subfigure}
\caption{\it 
Examples of source configurations for two-charge solutions.  
(Left) The black curve results from a macroscopic number of quanta in the lowest mode; it sources a geometry which is a spectral flow of the global $AdS_3\times\SS^3$ geometry.  On the other hand, putting a single quantum in the highest available mode (the tightly coiled spiral shown in red) makes an orbifold geometry $(AdS_3\times \SS^3)/\ZZ_{n_1n_5}$.
(Right) An NS5-P supertube separates the individual fivebranes onto the Coulomb branch by a fixed amount determined by the momentum charge.  
}
\label{fig:TwoExtremes}
\end{figure}

For instance, an F1-P source with $n_1n_5/k$ excitations of a given $k^{\rm th}$ oscillator mode dualizes to an $(AdS_3\times \SS^3)/\ZZ_k$ geometry.  At one extreme, $k=1$ yields global $AdS_3\times \SS^3$, spectrally flowed to the RR sector; the dual string source has the largest number of quanta excited ($n_1n_5$ excitations of the lowest mode) and the KK monopole loop has a radius of order the AdS scale.
At the other extreme, a single oscillator excitation of mode number $k=n_1n_5$ yields the deepest AdS throat, with the dual KK monopole loop executing a tight spiral $n_1n_5$ times along a circle whose radius is $n_1n_5$ times smaller than the AdS scale.  These two sources are depicted in figure~\ref{fig:TwoExtremes}.

A more tightly coiled source yields a deeper throat, much like putting fivebranes in a more closely spaced arrangement yields a deeper throat.  A meandering source locus, such as the black curve in figure~\ref{fig:TwoExtremes}, yields a geometry whose excitations are gapped at the AdS scale; on the other hand, the most tightly coiled sources, such as the red curve in figure~\ref{fig:TwoExtremes}, yield a geometry whose excitation gap is up to $n_1n_5$ times smaller than the AdS scale.  The excitation gap is a measure of the redshift to the bottom of the throat.

In two-charge geometries, the source configuration is frozen, specified by a discrete choice of an integer partition of $N=n_1n_5$; there is an equilibrium position that keeps the sources separated on the Coulomb branch, and it costs energy to push the source strands together.  One can engineer a similar situation in little string theory by considering an NS5-P supertube (the F1-P solution dualized slightly differently, see the right-hand plot in figure~\ref{fig:TwoExtremes}) that supports the fivebranes at a fixed radius determined by the (angular) momentum carried by the configuration.

To find an analogue of the Coulomb branch moduli space of little string theory, one must consider three-charge BPS solutions which have a deep throat scaling limit%
~\cite{Bena:2006kb}.  
It is this class of geometries which we now consider.


\section{Three-charge microstate geometries}
\label{sec:SugraFrames}

Three-charge geometries with angular momentum have been considered in a variety of duality frames, see for instance%
~\cite{Denef:2002ru,Bena:2005va,Berglund:2005vb,Bena:2011dd}.  
We will be interested in three such duality frames:
\begin{itemize}
\item
Type IIA on $\TT^6$ with D2-branes wrapping mutually orthogonal $\TT^2\subset \TT^6$
\be
\label{IIAcharge}
\begin{tabular}{| l | l |}
\hline 
Conserved Charge 
&
Dipole Charge
\\
\hline
D2:~$5\,6\,~\,~\,~\,~$
&
D4:~ $~\,~\,~7\,8\,9\,10$
\\
\hline
D2:~$~~\,~\,7\,8\,~~\,~$
&
D4:~ $5\,6\,~~\,~\,9\,10$
\\
\hline
D2:~$~~~\,~\,~\,~\,9\,10$
&
D4:~ $5\,6\,7\,8\,~~~~~$ 
\\
\hline
D0:
&
D6:~ $5\,6\,7\,8\,9\,10$ 
\\
\hline
\end{tabular}
\ee
\item
M-theory on $\TT^6$ with wrapped M2-branes, obtained by taking the strong coupling limit of the above IIA configuration; let $\psi$ be the coordinate of the M-theory circle
\be
\label{Mcharge}
\begin{tabular}{| l | l |}
\hline 
Conserved Charge 
&
Dipole Charge
\\
\hline
M2:~$5\,6\,~\,~\,~\,~$
&
M5:~ $~\,~\,~7\,8\,9\,10\,\psi$
\\
\hline
M2:~$~~\,~\,7\,8\,~~\,~$
&
M5:~ $5\,6\,~~\,~\,9\,10\,\psi$
\\
\hline
M2:~$~~~\,~\,~\,~\,9\,10$
&
M5:~ $5\,6\,7\,8\,~~~\,~~\psi$ 
\\
\hline
J:~$~~~~~~~\psi$
&
KKM:~ $1\,2\,3\,\psib$ 
\\
\hline
\end{tabular}
\ee
(we denote KK monopoles by their {\it transverse} dimensions, with the fibered circle in boldface).  Note that the D0 charge lifts to angular momentum, because the M-theory circle is contractible inside the KK monopole cores.
\item
Type IIB on $\TT^4\times \SS^1$ with D3-branes wrapping mutually orthogonal $\TT^2\subset \TT^4$ and intersecting over $\SS^1$, obtained by shrinking the torus wrapped by one of the M2-brane stacks to sub-Planckian size.  The shrinking $\TT^2$ dualizes to an $\SS^1$ in IIB which we parametrize by $v$.  This configuration is trivially T-dual to D1-D5.
\be
\label{IIBcharge}
\begin{tabular}{| l | l |}
\hline 
Conserved Charge 
&
Dipole Charge
\\
\hline
D3:~$5\,6\,~~~~ v$
&
D3:~ $~~~~7\,8\,\psi$
\\
\hline
D3:~$~~~~7\,8\,v$
&
D3:~ $5\,6\,~~~~\psi$
\\
\hline
P:~$~~~~~~~ v$
&
KKM:~ $1\,2\,3\,\vb$ 
\\
\hline
J:~$~~~~~~~\psi$
&
KKM:~ $1\,2\,3\,\psib$ 
\\
\hline
\end{tabular}
\ee
\end{itemize}
We now present some details of the geometries sourced by these objects in their respective duality frames.

\subsection{The standard 11d 3-charge BPS background}

We start with the standard 11d 3-charge background of Bena and Warner~\cite{Bena:2007kg}.  This background of M-theory compactified on $\TT^6 = \TT^2 \times \TT^2 \times \TT^2$ (along directions $56\;78\;9\underline {10}$) has M2-brane sources along each of the component $\TT^2$'s.  The metric is
\begin{equation} \label{11d metric}
\begin{split}
\dd s_{11}^2 &= - (Z_1 Z_2 Z_3)^{-2/3} \, (\dd t + k)^2 + (Z_1 Z_2 Z_3)^{1/3} \, \dd s_4^2 \\
&  + \bigg( \frac{Z_2 Z_3}{Z_1^2} \bigg)^{1/3} \, (\dd y_5^2 + \dd y_6^2) + \bigg( \frac{Z_1 Z_3}{Z_2^2} \bigg)^{1/3} \, (\dd y_7^2 + \dd y_8^2) + \bigg( \frac{Z_1 Z_2}{Z_3^2} \bigg)^{1/3} \, (\dd y_9^2 + \dd y_{10}^2)~,
\end{split}
\end{equation}
where the $Z_I$ are functions defined on a four-dimensional hyperK\"ahler base $\cB_4$ with metric $\dd s_4^2$, and $k$ is a 1-form having legs on $\dd s_4^2$.  The 3-form potential is given by
\begin{equation} \label{11d 3form}
\cA_3 = A^1 \wedge \dd y_5 \wedge \dd y_6 + A^2 \wedge \dd y_7 \wedge \dd y_8 + A^3 \wedge \dd y_9 \wedge \dd y_{10}~,
\end{equation}
where the 1-forms $A^I$ are given by
\begin{equation} \label{AI}
A^I = - Z_I^{-1} \, (\dd t + k) + \hat B^I~, \qquad \text{and we define} \qquad \Theta^I \equiv \dd \hat B^I~,
\end{equation}
where we use a hat on the 1-forms $\hat B^I$ to distinguish them from the Kalb-Ramond field~$B$.

The background defined by \eqref{11d metric} and \eqref{11d 3form} then gives a supersymmetric solution to 11d supergravity when the various quantities defined solve the BPS equations:
\begin{equation} \label{BPS eqns}
\begin{split}
\Theta^I &= \hodge_4 \Theta^I~, \\
\dd \hodge_4 \dd Z_I &= \frac12 C_{IJK} \, \Theta^J \wedge \Theta^K~, \\
\dd k + \hodge_4 \dd k &= Z_1 \, \Theta^1 + Z_2 \, \Theta^2 + Z_3 \, \Theta^3~,
\end{split}
\end{equation}
where the intersection numbers $C_{IJK} = \abs{\varepsilon_{IJK}}$.

\subsection{Dualizing to the IIB frame}\label{IIBgeom}

We want to dualize this M2-M2-M2 solution into IIB supergravity with charges D3-D3-P, so we first make a reduction to IIA with D2-D2-F1 background charges,%
\footnote{Note that this is different from the reduction that leads to the IIA frame~\pref{IIAcharge}.}
and then T-dualize along the~F1.
We leave the details for the various duality transformations to Appendix \ref{T duality conventions}.  In the IIB frame, we obtain the NS fields
\begin{align}
\begin{split} \label{IIB metric}
\dd s_\text{IIB}^2 &= - \frac{1}{Z_3 \sqrt{Z_1 Z_2}} \, (\dd t + k)^2 + \sqrt{Z_1 Z_2} \, \dd s_4^2 + \frac{Z_3}{\sqrt{Z_1 Z_2}} \big( \dd v +A^3 \big)^2 \\
& \qquad \qquad \qquad + \sqrt{\frac{Z_2}{Z_1}} \big( \dd y_5^2 + \dd y_6^2 \big) + \sqrt{\frac{Z_1}{Z_2}} \big( \dd y_7^2 + \dd y_8^2 \big)~,
\end{split} \\
e^\phi &= 1~, \\
\tilde B_2 &= 0~.
\end{align}
Since the Kalb-Ramond field now vanishes, we will use $B^I$ exclusively to refer to the three 1-forms appearing in the 1-form potentials
\begin{equation}
A^I = - Z_I^{-1} \, (\dd t + k) + B^I, \qquad \Theta^I \equiv \dd B^I~.
\end{equation}
The RR fields $G_1, G_3$ vanish, and $G_5$ is given by
\begin{equation}
G_5 = (1 + \hodge_{10}) \Big( \dd A^1 \wedge \dd y_5 \wedge \dd y_6 + \dd A^2 \wedge \dd y_7 \wedge \dd y_8 \Big) \wedge \big( \dd v + A^3 \big)~.
\end{equation}
This can be written out as
\begin{equation} \label{5form flux}
\begin{split}
G_5 &= \Big( \dd A^1 \wedge \dd y_5 \wedge \dd y_6 + \dd A^2 \wedge \dd y_7 \wedge \dd y_8 \Big) \wedge \big( \dd v + A^3 \big) \\
& \qquad - \Big[ Z_1^{-1} Z_3^{-1} \, (\dd t + k) \wedge \big( Z_2 \hodge_4 \Theta^2 - \hodge_4 \dd k \big) + \hodge_4 \dd Z_2 \Big] \wedge \dd y_5 \wedge \dd y_6 \\
& \qquad - \Big[ Z_2^{-1} Z_3^{-1} \, (\dd t + k) \wedge \big( Z_1 \hodge_4 \Theta^1 - \hodge_4 \dd k \big) + \hodge_4 \dd Z_1 \Big] \wedge \dd y_7 \wedge \dd y_8~.
\end{split}
\end{equation}
One has that $\dd G_5 = 0$ as a consequence of the BPS equations \eqref{BPS eqns}.  Thus one can integrate this to obtain a $C_4$ such that $G_5 = \dd C_4$.  There are many possible gauge choices, and we will find it convenient to choose
\begin{equation}
C_4 = \Big( y_5 \, \dd A^1 \wedge \dd y_6 + y_7 \, \dd A^2 \wedge \dd y_8 \Big) \wedge \big( \dd v + A^3 \big) + \cdots.
\end{equation}

\subsection{Gibbons-Hawking base}

The above geometries simplify when the four-dimensional hyperK\"ahler base $\cB_4$ is taken to be a Gibbons-Hawking space, a circle fibration over $\RR^3$
\be
\label{Gibbons-Hawkingmetric}
\dd s_4^2 = V^{-1}(d\psi+A)^2 + V \dd\yy\cdot \dd\yy \quad,\qquad \vec\nabla V = \vec\nabla\times\vec A 
~,
\ee
where $\psi$ parametrizes the circle.
The various metric coefficients $Z_I$, $k$, $V$, $A$ \etc. are then harmonic functions and forms on the $\RR^3$ parametrized by $\yy$.  The choice of harmonic function
\be
\label{Vfn}
V = \epsilon_0 + \sum_a \frac{q_a}{r_a} 
\quad , \qquad 
r_a = {|\yy-\yy_a|}
~,
\ee
specifies the locations $\yy_a$ of KK monopole cores.  The remaining harmonic functions/forms are
\bea
\label{NoSingCond}
Z_I &=& {\textstyle\half} (C_{IJK} K^J K^K)/V + L_I
\nonumber\\
k &=& \mu(d\psi+A) + \omega
\\
\mu &=& \coeff16 (C_{IJK} K^I K^J K^K)/V^2 + \coeff12 (K^I L_I)/V + M
\nonumber\\
\vec\nabla\times\vec\omega &=& V\vec\nabla M - M\vec\nabla V + \coeff12(K^I\vec\nabla L_I - L_I \vec\nabla K^I)
\nonumber
\eea
with $C_{IJK} = |\epsilon_{IJK} |$  the triple-intersection product, and
\bea
\label{KLdefs}
K^I = \kappa_0^I + \sum_{a=1}^N \frac{k_a^I}{r_a}
\quad ~&,& \qquad
\nonumber\\
L_I = \ell_{0,I} + \sum_{a=1}^N \frac{\ell_{a,I}}{r_a}
\quad &,& \qquad
\ell^a_I = -\coeff12 C_{IJK} \frac{k^J_a k^K_a}{q_a}
\\
M = m_0 + \sum_{a=1}^N \frac{m_a}{r_a}
\quad &,& \qquad
m_a = \coeff1{12} C_{IJK} \frac{k^I_a k^J_a k^K_a}{q_a^2} = \frac{k^1_a k^2_a k^3_a}{2 q_a^2}  ~.
\nonumber
\eea
This specification of the residues $\ell_{a,I}$ and $m_a$ in terms of the local dipole charges $k_a^I$ renders the supergravity solution regular up to orbifold singularities.
The asymptotically $AdS_3$ form of the metric determines
\begin{gather}
\label{netcharges}
\epsilon_0= 0
~~,~~~
q_0 = \sum_{a=1}^N q_a = 1
~~,~~~
\kappa_0^I = 0 
~~,~~~
\ell_{0,1} = 0 
~~,~~~
\ell_{0,2} = 0 \\
\ell_{0,3} = 1
~~,~~~
m_0 = -\coeff12 q_0^{-1} \sum_{a=1}^N k_a^3 ~.
\end{gather}

\begin{figure}
\centerline{\includegraphics[width=2.5in]{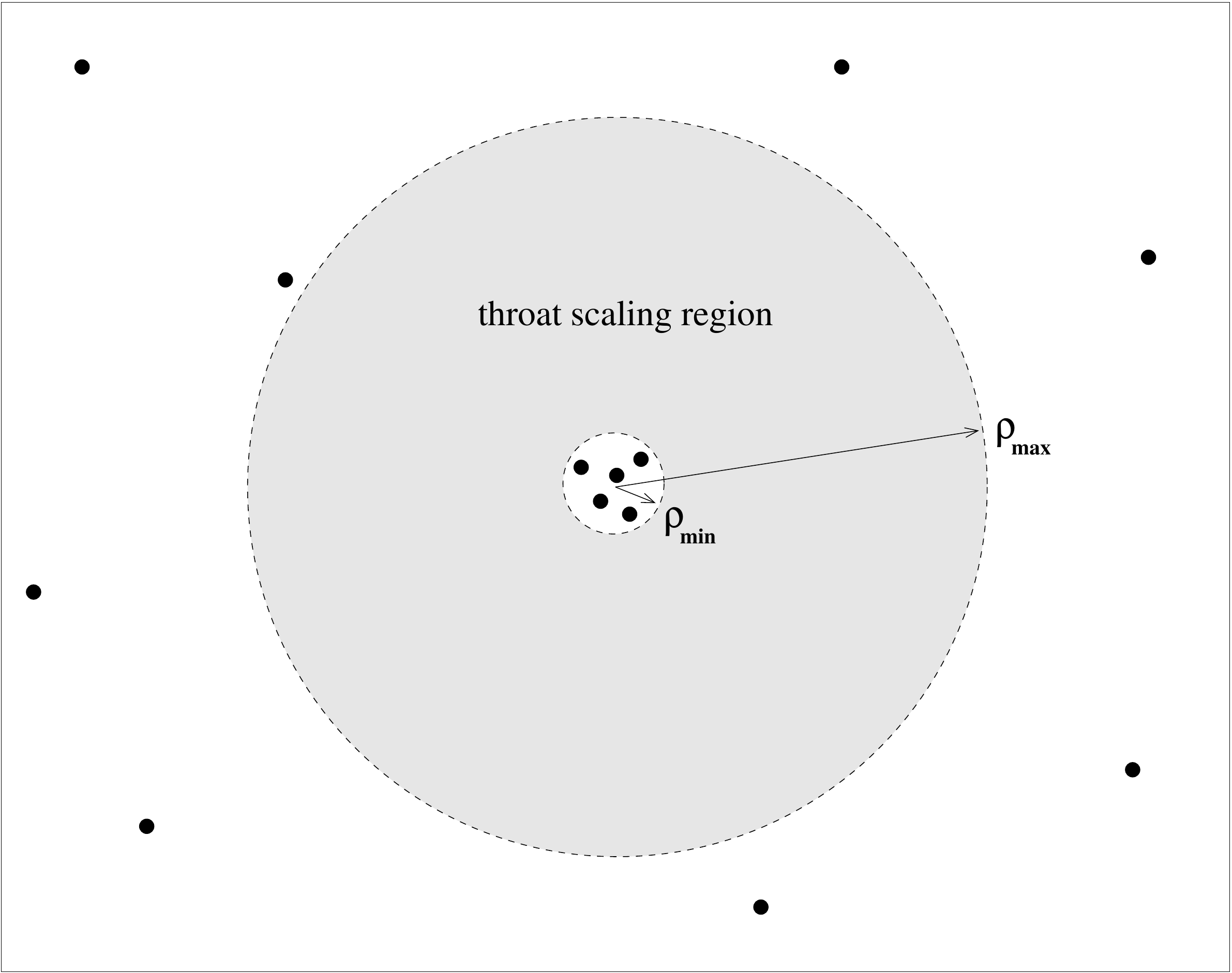}}
\setlength{\unitlength}{0.1\columnwidth}
\caption{\it 
Scaling together a cluster of poles leads to a throat whose depth is controlled by the size of the cluster $\rho_{\rm min}$ relative to its distance $\rho_{\rm max}$ to other poles.}
\label{fig:ThroatRegion}
\end{figure}

By Gauss' law, the net charge within a region will be governed by the total source strength inside the surface bounding the region; scaling together a cluster of charges as in figure~\ref{fig:ThroatRegion}, one finds a self-similar throat which is capped off at a depth controlled by the typical source separation within the cluster, much as in the NS5 throat of little string theory on its Coulomb branch.

\begin{figure}
\centerline{\includegraphics[width=4.0in]{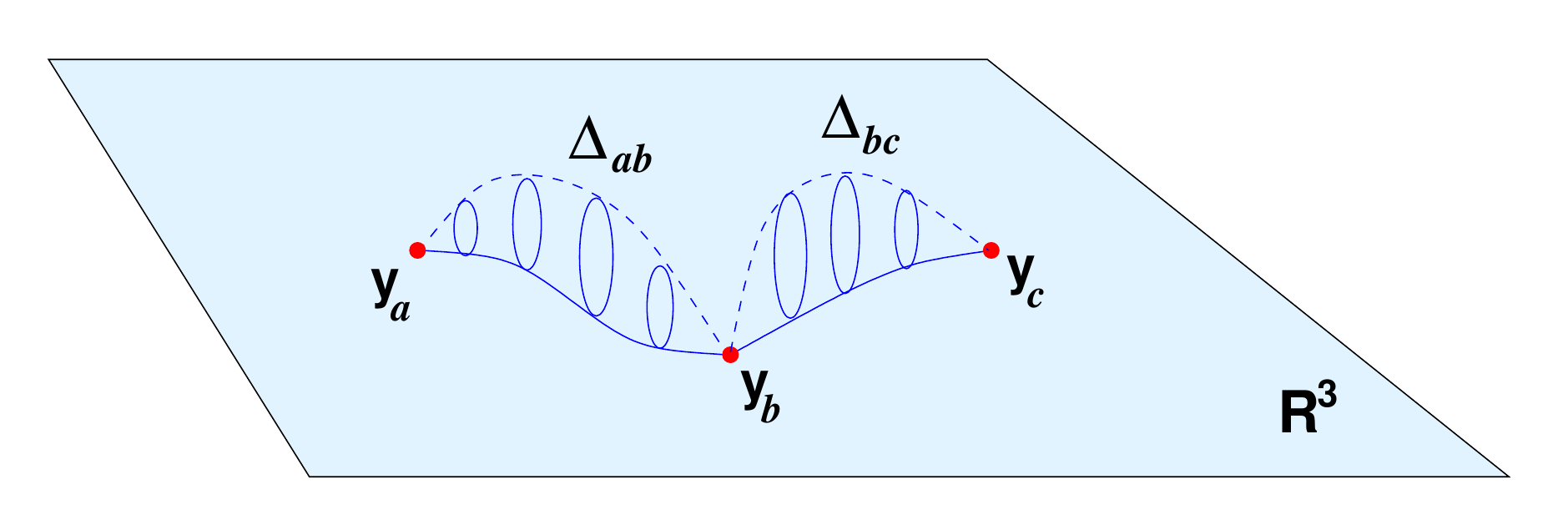}}
\setlength{\unitlength}{0.1\columnwidth}
\caption{\it 
Poles in the the metric function $V$ cause the degeneration of the $\SS^1_\psi$ fiber of the Gibbons-Hawking base $\cB_4$.  Each pair of points defines a two-cycle characterized by $\SS^1_\psi$ fibered over a path $\cC_{ab}$ between poles $\yy_a$ and $\yy_b$.}
\label{fig:Gibbons-HawkingBase}
\end{figure}

In little string theory, the objects becoming light near the origin of the Coulomb branch were the W-branes stretching between separated NS5's.  A similar structure holds in microstate geometries.  For every pair of poles $(a,b)$ in $V$, there is a a two-cycle $\Delta_{ab}$ in the Gibbons-Hawking base $\cB_4$, see figure~\ref{fig:Gibbons-HawkingBase}.  The analogue of the W-branes of little string theory are branes stretching over the cycle $\Delta_{ab}$.

The volume of the two-cycle $\Delta_{ab}$ in the Gibbons-Hawking metric~\pref{Gibbons-Hawkingmetric} on $\cB_4$ is $4\pi|\yy_b-\yy_a|$.  Similarly, all the factors of $Z_I$, $V$, \etc. drop out of the determinant of the induced metric on the worldvolume of a probe M2-brane wrapping this cycle in the M-theory duality frame~\pref{Mcharge}, or of a probe D3-brane wrapping the dual three-cycle $\widetilde\Delta_{ab}$ in the type IIB duality frame~\pref{IIBcharge}.  The mass of such wrapped branes formally vanishes in the limit $|\yy_b-\yy_a|\to 0$.%
\footnote{More precisely, the cycle has a finite proper spatial volume, but redshifts away down a long throat so that the energy cost of a wrapped brane vanishes in the limit.  Thus the effect is the same as if the cycle were shrinking away.}
Thus, scaling down the size of a cluster of poles in the harmonic functions shrinks the associated two- or three-cycles, and one expects a phase transition to the Higgs branch of condensed W-branes in the scaling limit.

\section{Geometries and quivers}
\label{sec:QuiverQM}

As discussed above, the CFT dual is ill-suited for answering questions about local geometry on the gravity side.  However, the microstate geometry solutions have collective coordinates in the pole locations $\yy_a$ which describe the location of background charge sources, and one may consider the state space that results from quantizing these collective coordinates.  This is expected to provide a useful approximation in situations with enough supersymmetry, provided that these collective coordinates are the lightest degrees of freedom in the system and there is a gap to exciting other modes. 

The collective coordinates are subject to a set of constraints known as the {\it bubble equations}
\be
\label{bubbleqs}
\sum_{b\ne a}\frac{\langle\Gamma_a,\Gamma_b\rangle}{|\yy_a - \yy_b|} = \langle\Phi_0,\Gamma_a\rangle
\ee
where $\Gamma_a$ is the eight-vector of charges~\pref{IIAcharge}-\pref{IIBcharge}, and $\Phi_0$ is the harmonic potential background
\be
\Gamma_a = (q_a, \ell_I^a, k_a^I, m_a)
\quad,\qquad
\Phi \equiv (\epsilon_0,\ell_I^0,\kappa_0^I,m_0) ~;
\ee
the symplectic inner product $\langle*,*\rangle$ is
\be
\label{symprod}
\Gamma_{ab} \equiv {\langle\Gamma_a ,\Gamma_b\rangle} = 2(q_b m_a - q_b m_a) + \sum_{I=1}^3 (\ell_I^b k_a^I - k_b^I\ell^a_I ) ~.
\ee
The bubble equations place $N-1$ constraints on the $3(N-1)$ parameters $\yy_a$ (modulo translations); these constraints depend parametrically on the data $q_a$, $k_a^I$ specifying the charges.  

The collective coordinates $\yy_a$ are scalars in a collection of vector supermultiplets $Y_a$.
The collective coordinate approximation fails when the W-branes start to become light, since the collective coordinates are then not the only light degrees of freedom in the effective dynamics of the charge centers.  As a crude approximation to the incipient Higgs branch dynamics, one might include the lightest stretched W-branes as well in the effective theory; this is the approach of {\it quiver quantum mechanics}.  These W-branes are the quanta of bifundamental hypermultiplets $\Phi_{ab}$ in the quiver.
An encouraging result in this regard%
~\cite{Denef:2002ru} 
is that the bubble equations~\pref{bubbleqs} arise as the D-term equations for the vector multiplets of the quiver QM upon integrating out the hypermultiplets, which is a valid approximation well out on the Coulomb branch.

In the IIA duality frame~\pref{IIAcharge}, the hypermultiplet quanta are fundamental strings stretching between primitive (entropy-less) {D6-D4-D2-D0} bound states, with mass ${|\yy_b-\yy_a|}/\alpha' $.  These lift to M2-branes stretching over the cycles $\Delta_{ab}$ in the M-theory frame~\pref{Mcharge}, which then descend to D3-branes stretching over three-cycles $\widetilde\Delta_{ab}$ which are roughly $\Delta_{ab}\times \SS^1_v$ in the type IIB frame~\pref{IIBcharge}.  More precisely, the three-cycles in type IIB are $\TT^2$ fibrations over a path $\cC_{ab}$ between $\yy_a$ and $\yy_b$ whose detailed structure will be analyzed below.  
There is enough supersymmetry to guarantee that the mass of the lightest stretched brane is the same in all duality frames, in particular it is proportional to $|\yy_b-\yy_a|$.

With 8 or 16 supersymmetries one sometimes captures the full dynamics of supergravity in a decoupling limit of supersymmetric QM; these are the situations where Matrix Theory applies (for a review, see for instance%
~\cite{Banks:1999az}).  In the present case, the background preserves only four supersymmetries; it is too much to ask that the whole spacetime be captured by the quiver dynamics.  But that is not how we want to use the quiver here; rather, the quiver encodes simply the collective coordinate dynamics in a much larger system, similar to the way one quantizes the low-energy dynamics of monopoles and dyons in nonabelian gauge theory (for a review, see%
~\cite{Weinberg:2006rq}).  And like the collective coordinate quantization of monopoles, the states arising from quiver QM are not guaranteed to be a complete description of the low-energy structure; in the monopole problem, nonabelian degrees of freedom in the monopole cores become important when they collide, similarly there are degrees of freedom left out of the quiver QM when charge centers coalesce.  Nevertheless one may entertain the possibility that the QM state space captures some fraction of the BPS spectrum, and investigate how large a fraction that might be.  

How much entropy is one looking to find?  The BPS gravitational entropy is determined by the conserved charges and dipole charges of the supergravity solution.
These charges are determined by the data $q_a$, $k_a^I$ which specify the supergravity solution~\pref{11d metric} or~\pref{IIB metric} via~\pref{NoSingCond}-\pref{KLdefs}, and receive a contribution from each pole location in the harmonic functions, beginning with the dipole charge
\bea
d^I_a &=&  2 \Bigl(k^I_a - q_a  \sum_b k^I_b \Bigr)
\nonumber\\
d^I &=& \sum_a d^I_a 
~.
\eea
The regularity conditions~\pref{KLdefs} then determine the conserved charges in terms of the dipole charges
\bea
Q_I &=& \sum_a Q_{aI} = \frac12 C_{IJK} \,\sum_a d_a^J d_a^K / q_a
\nn\\
J_{\psi} &=& \sum_a J_{a\,\psi} = \frac16 C_{IJK} \sum_a d_a^I d_a^J d_a^K /q_a^2
~.
\eea
The dipole charges and pole locations also determine the transverse angular momentum
\be
\label{Jperp}
J_T = 4\Bigl| \sum_{a,I} d^I_a \,\yy_a \Bigr|
~.
\ee

The black hole/black ring entropy with these charges is given by 
\be
\label{ringent}
S^{~}_{\it \! BH} = \pi \sqrt{\cI_4} ~,
\ee
in terms of the quartic $E_{7(7)}$ invariant
\begin{align}
\label{E77Charge}
\cI_4 = & 2( d^1 d^2 Q_1 Q_2 + d^2 d^3 Q_2 Q_3 + d^3 d^1 Q_3 Q_1) 
- (d^1 Q_1)^2 - (d^2 Q_2)^2 - (d^3 Q_3)^2 
\nn\\
& \qquad - d^1 d^2 d^3 [4 J_L + 2(d^1 Q_1 + d^2 Q_2 + d^3 Q_3) - 3 d^1 d^2 d^3]
~.
\end{align}

We wish to compare this gravitational entropy with the number of BPS states of quiver QM.
Scaling down the size of a cluster of charge centers $S$, the geometry develops a horizon.  At the same time, the hypermultiplets of the corresponding quiver QM become effectively massless and condense, and so the horizon dynamics involves a {\it wrapped brane condensate}.  The resulting states are often denoted as `pure-Higgs' states, to distinguish them from nearby states on the Coulomb branch, which interestingly also have a footprint on the Higgs branch which can be computed by integrating out the vector multiplets and studying the resulting effective dynamics of the hypermultiplets%
~\cite{Denef:2002ru,Sen:2011aa,Bena:2012hf}. 

The approach of truncating the full supergravity dynamics to the quiver degrees of freedom is not expected to fully capture the horizon dynamics; the horizon that forms should not carry details of how it was assembled, \eg\ the data of the specific quiver used.  
For instance, nonabelian degrees of freedom that redistribute charges and fluxes between the charge centers%
~\cite{Polchinski:1997pz}
are missing.
Nevertheless, one may hope that this approach captures much of the essential physics.

In the IIA picture~\pref{IIAcharge}, consider a set of primitive bound states with $\it (D6,D4,D2,D0)$ brane charges
\be
\Gamma_a = (q_a, k_a^I, \ell_{Ia}, m_a)
\ee
for centers $a=1\ldots N$.  The position of each center in $\RR^3$ is specified by a vector multiplet $Y_a$, leading to $U(1)^N$ gauge QM.%
\footnote{We will ignore the possibility of multiple charge centers having identical charge assignments, which would lead to a nonabelian gauge theory and the associated identical particle statistics on the Coulomb branch.}  
In addition, for each pair of centers there are 
$\Gamma_{ab} = \langle\Gamma_a,\Gamma_b\rangle$
hypermultiplets $\Phi_{ab}^{(\alpha)}$, $\alpha=1\ldots \Gamma_{ab}$, on the links of the quiver (the sign of $\Gamma_{ab}$ gives an orientation to the links).  The quantity $\Gamma_{ab}$ is the symplectic inner product of the charge vectors, and can be thought of as the number of mutual intersections of the primitive brane bound states on $\TT^6$ when they coincide in $\RR^3$.  There are similar explanations for the number $\Gamma_{ab}$ of links on the quiver in the M-theory and type IIB duality frames; for the moment, let us continue with the comparison of the entropy exhibited by the quiver construction with that of the corresponding black hole.  We will derive the number $\Gamma_{ab}$ from the M-theory and type IIB perspectives in the next section.

The bubble equations~\pref{bubbleqs} can obstruct the existence of a scaling solution where charge centers can coalesce at no cost in energy.  For instance, the cluster of charge centers has to be able to shed its contribution to the angular momentum $J_T$, equation~\pref{Jperp}, apart from the overall contribution of the cluster center of mass; this requirement forces the cluster to consist of more than two centers. 
When the quiver has an oriented closed loop, the bubble equations admit scaling solutions, where the collection $S$ of nodes along the loop corresponds to objects whose locations $\yy_a$ can be scaled together to make an arbitrarily deep throat%
~\cite{Denef:2002ru}.
The simplest example is the triangular quiver, where the conditions for a scaling solution are the satisfaction of the triangle inequalities
\be
|\Gamma_{12}| + |\Gamma_{23}| \ge |\Gamma_{31}|
\ee
(and cyclic permutations).  

The entropy of the three-node quiver describing a three-center scaling cluster has been computed in~\cite{Denef:2007vg,Bena:2012hf}, with the result
\be
\label{Striangle}
S^{~}_\triangle \sim n_{12}\log(2n_{12}) - (n_{23}+n_{31}-n_{12})\log (n_{23}+n_{31}-n_{12}) + {\rm cyclic}
~,
\ee
where $n_{ab}=|\Gamma_{ab}|$.
The leading asymptotic $n\log n$ behavior in this expression cancels among the various terms, resulting in a leading scaling behavior linear in $n_{ab}$.

Note that the conserved charges $Q_{Ia}\propto C_{IJK} d^J_a d^K_a$ are quadratic in the dipole charges $d^J_a$, and the symplectic product $\Gamma_{ab}$, equation~\pref{symprod}, pairs conserved charges with dipole charges.  Therefore, when we scale all the $d^I_a$ by a common factor $\lambda_I$, the entropy of pure-Higgs states will scale as $\lambda_1\lambda_2\lambda_3$, which is exactly the same way that the black ring entropy~\pref{ringent} scales when the charges $Q_I$ are scaled by the corresponding factor $C_{IJK}\lambda_J\lambda_K$.  This provides encouragement that the degrees of freedom kept in the truncation to quiver QM are some finite fraction of the degrees of freedom which account for the black hole/black ring entropy.  

As a quantitative example, consider the simplest three-node scaling cluster with the choice of charges analyzed in%
~\cite{Bena:2007qc}.  
The supergravity solution has four charge centers, with charges
\begin{align}
&q_a = (210,-105,-105,1)
&k_a^2 = (-20000,16000,7887,0)
\nn\\
\label{BWWcharges}
&k_a^1 = (525,1200,2210,0)
&k_a^3 = (6400,1613,7900,0)~~~~\,.
\end{align}
A fourth point must be added so that when scaling the cluster of the first three, the transverse angular momentum is conserved and all the bubble equations are solved.

The Gibbons-Hawking base has four charge centers, with 
\be
V = \frac{2Q}{|\yy-\yy_1|} - \frac{Q}{|\yy-\yy_2|} - \frac{Q}{|\yy-\yy_3|} + \frac{1}{|\yy-\yy_4|}
~,
\ee
where $Q=105$.  The scaling limit takes the first three centers to a common point, which remains a finite distance $\rho$ from the last point which we take to be at $\yy=0$.  

Plugging in the charge assignments~\pref{BWWcharges}, one finds
\be
S^{~}_\triangle / S_{\rm BH} \approx 0.04
~.
\ee
This result is quite intriguing.  It suggests that the horizon of the BPS black hole in the M-theory duality frame is coated with a condensate of M2-branes that wrap the cycles $\Delta_{ab}$, reminiscent of the analysis of~\cite{Gaiotto:2004pc,Denef:2007yt}, though the details here are rather different.%
\footnote{In~\cite{Gaiotto:2004pc,Denef:2007yt}, one was looking for the distinct BPS configurations of finite mass branes to account for the entropy; here the branes condense, and we are looking for the number of supersymmetric ground states that differ by the expectation values of the condensates (the distinct solutions to the D-term and F-term equations of the quiver QM).}
Similarly, in the type IIB frame, it seems that D3-branes wrap the horizon.

Thus, the ``hair-brane idea'': {\it A proper treatment of the assembly of black holes from the Coulomb branch of microstate geometries will exhibit the long string that carries the entropy of $AdS_3$ black holes as a collective excitation of the branes that become light at the entrance to the Higgs branch.}  Note that in the IIB frame, the stretched branes are D3-branes wrapping cycles which include $\SS^1_v$, and are thus ideal candidates to wind into an effective long string-like object as $\widetilde\Delta_{ab}$ shrinks and these branes become light.

These stretched branes are direct analogues of the stretched branes on the Coulomb branch of little string theory.  There, the little string appears when these objects condense (rather directly in IIA, or as a soliton in IIB).  KK monopoles are T-dual to NS5-branes%
~\cite{Gregory:1997te}, 
so in microstate geometries one expects ``W-string'' objects to appear in IIB, and to be somewhat hidden in IIA/M-theory frames, rather than the other way around.


\section{Probe brane dynamics}
\label{sec:probes}

A crucial ingredient of the quiver entropy count is the large number $\Gamma_{ab}$ of hypermultiplets on the links of the quiver.  In the IIA duality frame, one can understand this count in terms of the number of mutual intersections of the primitive brane bound states.  Alternatively, the lower-dimensional branes are dissolved as electric and magnetic fluxes of the D6-branes' worldvolume gauge fields, and so the endpoints of open strings stretching between primitive brane bound states are effectively electric charges in a strong magnetic field; $\Gamma_{ab}$ is simply the number of states in the lowest Landau level.  It is this latter explanation of the hypermultiplet accounting that lifts cleanly to the M-theory duality frame.

\subsection{M2-branes}\label{sec:M2}

Consider the M2-brane effective action
\begin{equation} \label{M2 action}
\cS_{M2} = - \tau_{M2} \int \vol_{M2} + \tau_{M2} \int \cA_3~.
\end{equation}
which pulls back the ambient metric (from which $\vol_{M2}$ is constructed) and three-form gauge field to the M2-brane worldvolume.  We wish to investigate the center-of-mass motion of an M2-brane wrapped on $\Delta_{ab}$ of the Gibbons-Hawking base and moving in the $\TT^6$ spanned by $y_{5-10}$.


First we pull back the metric~\eqref{11d metric} to obtain $\dd s_{M2}^2 \equiv i^* \dd s_{11}^2$.  Choose the $z$ axis in $\RR^3$ to run between the Gibbons-Hawking points that define this cycle.%
\footnote{For simplicity, we assume that the background is axisymmetric about the $z$ axis in $\RR^3$.  We expect that the essential aspects of the physics are more or less the same for non-axisymmetric backgrounds.}%
~Then the M2 extends along $t, \psi, z$, so we can use static gauge for these coordinates.  We assume that the M2 has some velocity in the $\TT^6$ directions, and hence $y_{5-10}$ are functions of $t$ (only).  The pullback metric is then
\begin{equation}
\label{M2 pullback metric}
\begin{split}
\dd s_{M2}^2 &= - (Z_1 Z_2 Z_3)^{1/3} \Big( (Z_1 Z_2 Z_3)^{-1} - \sum_I Z_I^{-1} (\frv_I)^2 \Big) \, \dd t^2 - \frac{2 \mu}{(Z_1 Z_2 Z_3)^{2/3}} \, \dd t \, \dd \psi \\
& \qquad  + \frac{\cQ}{(Z_1 Z_2 Z_3)^{2/3} V^2} \, \dd \psi^2 + (Z_1 Z_2 Z_3)^{1/3} \, V \, \dd z^2
~,
\end{split}
\end{equation}
where the $\frv_I$ are the velocities on each of the three $\TT^2$'s:
\begin{equation} \label{wv vel}
\frv_1 \equiv (\dot y_5, \dot y_6)
~, \qquad \frv_2 \equiv (\dot y_7, \dot y_8)
~, \qquad \frv_3 \equiv (\dot y_9, \dot y_{10})~.
\end{equation}
The determinant of the above metric is simply
\begin{equation}
- \det g_{M2} = 1 - V^{-1} \cQ \sum_I Z_I^{-1} (\frv_I)^2
\quad, \qquad 
\cQ \equiv Z_1 Z_2 Z_3 V - \mu^2 V^2~;
\end{equation}
here $\cQ$ is the $E_{7(7)}$ quartic invariant built out of the harmonic functions, which can be expanded in a form related to~\pref{E77Charge} as
\begin{equation}
\begin{split}
\cQ &= 
K^1 L_1 K^2 L_2 + K^1 L_1 K^3 L_3 + K^2 L_2 K^3 L_3 - \frac14 (K^I L_I)^2  \\
& \qquad \qquad  - 2 K^1 K^2 K^3 M - V^2 M^2 - M V (K^I L_I)  + V L_1 L_2 L_3 
~.
\end{split}
\end{equation}

For the 3-form potential $\cA_3$, 
it is simpler to choose a gauge for $\cA_3$ other than~\eqref{11d 3form},
\begin{equation} \label{11d 3form gauge}
\cA_3 = y_5 \, \dd A^1 \wedge \dd y_6 + y_7 \, \dd A^2 \wedge \dd y_8 + y_9 \, \dd A^3 \wedge \dd y_{10}~;
\end{equation}
then the pulled-back 3-form potential can be written
\begin{equation}
i^* \cA_3 = \dd t \wedge \Big( \sum_I F^I \, (\fra_I \cdot \frv_I) \Big)~, \qquad F^I \equiv \dd A^I~,
\end{equation}
where the $\fra_I$ are magnetic vector potentials:
\begin{equation} \label{wv vec pot}
\fra_1 \equiv (0, y_5)~, \qquad \fra_2 \equiv (0, y_7)~, \qquad \fra_3 \equiv (0, y_9)~,
\end{equation}
and hence one has terms with the products $(y_5 \dot y_6)$, $(y_7 \dot y_8)$, and $(y_9 \dot y_{10})$.

The full M2 brane action is then given by
\begin{equation}
\begin{split}
\cS_{M2} &= - \tau_{M2} \int_\cW \Big( 1 - V^{-1} \cQ \sum_I Z_I^{-1} (\frv_I)^2 \Big)^{1/2} \, \dd t \wedge \dd \psi \wedge \dd z \\
& \qquad \qquad + \tau_{M2} \int_\cW \dd t \wedge \Big( \sum_I F^I \, (\fra_I \cdot \frv_I) \Big)~,
\end{split}
\end{equation}
where $\cW \equiv \RR_t \times \Delta_{ab}$ is the worldvolume.  Since $\frv_I, \fra_I$ do not depend on $z$ and $\psi$, the $z$ and $\psi$ integrals can be done, in principle, leaving us an action for a relativistic, massive, charged particle moving in $\TT^6$ with velocities $\frv_I$, in a magnetic field given by the vector potentials $\fra_I$.

In practice, we can easily integrate $\psi, z$ in the WZ term.  We have
\begin{equation}
i^* \dd t \wedge F^I = - \dd t \wedge \dd \bigg( \frac{\mu}{Z_I} \bigg) \wedge \dd \psi + \dd t \wedge \Theta^I~.
\end{equation}
Since $\mu = 0$ at each Gibbons-Hawking point, the integral of the first term over $z$ vanishes.  The integral over $\psi, z$ in the second term gives $4 \pi \, \Pi^{(I)}_{ab}$.  Similarly, the integral over $\psi$ in the DBI term gives a factor of $4\pi$, so we can write the action as
\begin{equation} \label{M2 action 2}
{\cS}_{M2} = 4\pi  \tau_{M2} \int \dd t \bigg[ - \int_{z_a}^{z_b} \dd z \, \Big( 1 - V^{-1} \cQ \sum_I Z_I^{-1} (\frv_I)^2 \Big)^{1/2} + \sum_I \Pi^{(I)}_{ab} \, (\fra_I \cdot \frv_I) \bigg]~.
\end{equation}
Again, given the details of the solution, one can in principle do the integral in the kinetic term, since the $y_i$ do not depend on $z$.

It will be useful to write down the equations of motion for $y_9, y_{10}$ obtained from this action, since these coordinates are the ones that are dualized away in the IIB frame.  For $y_{10}$, we obtain the equation of motion
\begin{equation} \label{y10 eqn}
0 = \int_{z_a}^{z_b} \dd z \, \bigg( \frac{\cQ}{V Z_3} \bigg) \, \frac{\partial}{\partial t} \bigg[ \dot y_{10} \, \Big( 1 - V^{-1} \cQ \sum_I Z_I^{-1} (\frv_I)^2 \Big)^{1/2} \bigg] + \Pi^{(3)}_{ab} \, \dot y_9~,
\end{equation}
and for $y_9$,
\begin{equation} \label{y9 eqn}
0 = \int_{z_a}^{z_b} \dd z \, \bigg( \frac{\cQ}{V Z_3} \bigg) \, \frac{\partial}{\partial t} \bigg[ \dot y_{9} \, \Big( 1 - V^{-1} \cQ \sum_I Z_I^{-1} (\frv_I)^2 \Big)^{1/2} \bigg] - \Pi^{(3)}_{ab} \, \dot y_{10}~,
\end{equation}
with similar equations for the other two pairs of $y_i$.  For small velocities, the $\dot y_i^2$ terms may be neglected, and one obtains
\begin{equation}
\mathfrak{m} \, \ddot y_{10} + \Pi^{(3)}_{ab} \, \dot y_9 = 0, \qquad \mathfrak{m} \, \ddot y_9 - \Pi^{(3)}_{ab} \, \dot y_{10} = 0,
\end{equation}
which are the equations of motion for a charged particle of mass
\begin{equation} \label{mass}
\mathfrak{m} = \int_{z_a}^{z_b} \frac{\cQ}{V Z_3} \, \dd z
\end{equation}
moving in a constant, transverse magnetic field of strength $\Pi^{(3)}_{ab}$.%
\footnote{We note that subject to the regularity conditions, both $\cQ$ and $V Z_3$ should go as $1/r$ near the Gibbons-Hawking points, so the integrand in \eqref{mass} is everywhere finite. }


Thus, for M2-branes wrapping the cycles $\Delta_{ab}$, the magnetic four-form flux is experienced by the center-of-mass dynamics on $\TT^6$ as an effective two-form magnetic field on $\TT^2_{(I)}\subset \TT^6$ of strength%
~\cite{Gaiotto:2004pc,Denef:2007yt}
\be
\label{Pi flux}
\Pi_{ab}^{(I)} = \frac{1}{4\pi}\int_{\Delta_{ab}} \!\! \Theta^{I} =\left( \frac{k^I_b}{q_b} - \frac{k^I_a}{q_a} \right)
~,
\ee
and $\Gamma_{ab}$ is given by the product of fluxes through the two-cycle $\Delta_{ab}$%
~\cite{Bena:2005va,Berglund:2005vb},
\be
\label{hypernumber}
\Gamma_{ab} =  q_a q_b \Pi_{ab}^{(1)} \Pi_{ab}^{(2)}  \Pi_{ab}^{(3)} 
~,
\ee
which is the number of states in the lowest Landau level.
The additional factors of $q_{a,b}$ arise from the $\ZZ_q$ orbifold singularities at the poles of the cycle $\Delta_{ab}$; there are $q_{a,b}-1$ vanishing two-cycles at the orbifold locus, and the probe brane can wrap any of these as well as the two-cycle that stretches between the two poles, leading to $q_a q_b$ independent membrane states for each choice of Landau level state.  In this way, M-theory reproduces the multiplicity $q_a q_b$ of (Chan-Paton) labels for open strings stretching between stacks of $q_a$ and $q_b$ D6-branes in the type IIA reduction%
~\cite{Sen:1997kz}.

In the IIA D-brane collective dynamics, $\Gamma_{ab}$ is the number of species of open string stretching between the corresponding composite branes; there is one available open string state for each link on the quiver.  The explanation of this number as the degeneracy of the lowest Landau available to open strings in a D-brane magnetic field, lifts to an M2-brane wrapping a flux background in M-theory.  The D6 branes lift to KK monopoles in M-theory, the lower-dimensional branes are carried as magnetic $\cG_4$ flux, and the open strings have lifted to M2 branes wrapping the cycles $\Delta_{ab}$.  The intersection number is still the degeneracy of the lowest Landau level that the M2 branes occupy, and so once again is the number of states available to the lightest W-branes.  We see that all the ingredients of quiver QM have counterparts in the M-theory frame, and so the quiver dynamics is a truncation of the M-theory dynamics.  

One might worry that the M2-brane action becomes degenerate when $|\yy_b-\yy_a|\to 0$ and the volume term in~\pref{M2 action}, the usual source of kinetic energy, naively vanishes.  
Let $\cal S$ be a scaling cluster of Gibbons-Hawking points, whose center we place at the origin; denote the positions of these points
\be
\yy_a = \lambda\, \yscale_a
\quad,\qquad a\in \cal S~,
\ee
where $\yscale_a$ become fixed in the limit. In the scaling limit 
\be
\cQ \sim \lambda^{-4} 
~;~~~
V\, ,~ Z_I\, ,~\mu \sim \lambda^{-1} ~,
\ee
and so the terms in the induced metric~\eqref{M2 pullback metric} schematically scale as
\be
\dd s^2_{M2} \sim -\big(\lambda^{2} \,f_1 - \lambda^0 \,f_2\cdot(\frv_I)^2 \big)\, \dd t^2 - \lambda^{1}\, f_3\,\dd t\, \dd \psi
+ \lambda^0 \, \big[f_4\,\dd \psi^2 + f_5\,\dd \zscale^2 \big]
\ee
where the $f_i$ are functions of $\yscale$.
Thus the proper spatial size of a brane wrapping a scaling cluster remains constant, and dynamics is invariant if we also scale 
\be
t = \lambda^{-1} \, \hat t~,
\ee
in other words the dynamics redshifts commensurate with the depth of the throat, but does not degenerate.

The role of excitations above these ground states in the effective dynamics of the probe brane, particularly in the scaling limit, is an interesting question which we leave to future research.


\subsection{D3-branes}

In the type IIB duality frame, the origin of $\Gamma_{ab}$ lies in part in the structure of Landau levels available to a probe D3-brane wrapping the three-cycles $\widetilde\Delta_{ab}$ (dual to the two-cycles $\Delta_{ab}$ of the M-theory frame), and partly in the ground state structure of its worldvolume gauge field.  M-theory/type IIB duality involves singling out one of the mutually orthogonal $\TT^2$'s carrying the background M2-brane charge (say the $y_9$-$y_{10}$ directions) and shrinking it to sub-Planckian size.  In the process, the M-theory $\cG_4$ fluxes along $\Delta_{ab}$ times either $\TT^2_{(56)}$ or $\TT^2_{(78)}$ become five-form fluxes $G_5$ along these two-tori times $\widetilde\Delta_{ab}$; the effective center-of-mass motion within $\TT^4_{(5678)}$ of a D3-brane probe wrapping $\widetilde\Delta_{ab}$ is much the same as for an M2-brane probe wrapping $\Delta_{ab}$ in the original M-theory description.  The fate of the third four-form flux along $\Xi = \Delta_{ab}\times\TT^2_{(9\,10)}$ is more subtle, however.  The torus $\TT^2_{(9\,10)}$ dualizes into a circle $\SS^1_v$ in the type IIB frame, and the four-form flux along $\Xi$ dualizes into the structure of the fibration of this circle over the Gibbons-Hawking base and its interplay with $\Delta_{ab}$.  In this subsection, we delve into this structure.  To begin, we lay out our conventions for the D3-brane effective action.

\subsubsection{The D3 effective action}

The action of a D3-brane in a IIB SUGRA background is given by DBI and Wess-Zumino terms:
\begin{equation}
\cS_{\text{D3}} = \cS_{\text{DBI}} + \cS_{\text{WZ}}~,
\end{equation}
where, using $\cF \equiv 2 \pi \alpha' F$,
\begin{align}
\cS_{\text{DBI}} &=  - \tau_{D3} \int_{D3} \dd^4 \xi \, e^{-\phi} \, \sqrt{ - \det{(g + B + \cF)}}~, \\
\cS_{\text{WZ}} &= \tau_{D3} \int_{D3} \sum_{p} C_p \wedge e^{B + \cF} \wedge \sqrt{\frac{\hat \cA (4 \pi^2 \cR_T)}{\hat \cA (4 \pi^2 \cR_N)}}~.
\end{align}
Here $\tau_{D3}$ is the D3-brane tension, $\xi$ are some worldvolume coordinates, $\phi$ is the dilaton, $g, B$ are the (pullbacks of) the metric and NS 2-form potential, and $\cF$ is the 2-form field strength living on the brane which is sourced by the endpoints of open strings.  $C_p$ are the RR $p$-form potentials, $\hat \cA$ is the Dirac ``A-roof" genus, $\cR_T, \cR_N$ are the curvature 2-forms of the tangent and normal bundles, respectively.  For D3-branes, the $\hat \cA$ part can be expanded in a sum which terminates at 4-forms:
\begin{equation}
\sqrt{\frac{\hat \cA (4 \pi^2 \cR_T)}{\hat \cA (4 \pi^2 \cR_N)}} = 1 - \frac{1}{48} p_1(4 \pi^2 \cR_T) + \frac{1}{48} p_1(4 \pi^2 \cR_N)~,
\end{equation}
where $p_1$ is the first Pontrjagin class.  Therefore the WZ action can be expanded
\begin{equation}
\begin{split}
\cS_{\text{WZ}} &= \tau_{D3} \int_{\text{D3}} \bigg[ C_4 + C_2 \wedge (B + \cF) \\
& \qquad \qquad + C_0 \bigg( \frac12 (B + \cF) \wedge (B + \cF)  - \frac{1}{48} p_1(4 \pi^2 \cR_T) + \frac{1}{48} p_1(4 \pi^2 \cR_N) \bigg) \bigg]~,
\end{split}
\end{equation}
for the IIB RR potentials $C_4, C_2, C_0$.  As we saw in section~\ref{IIBgeom}, the potentials $C_2$, $C_0$, as well as the NS-NS $B$-field all vanish in our case, so the WZ action reduces to the $C_4$ term only.

The DBI part can also be expanded, defining $b \equiv B + \cF$, as follows:
\begin{equation}
\label{alt DBI}
\cS_\text{DBI} = - \tau_{D3} \int_{D3} e^{-\phi} \Big( 1 + \norm{b}^2 + \frac14 \norm{b \wedge b}^2 \Big)^{1/2} \, \vol_{D3}~,
\end{equation}
where the squared norm $\norm{\omega_{(p)}}^2$ of a $p$-form $\omega_{(p)}$ defined by
\begin{gather}
\omega_{(p)} \wedge \hodge_{D3} \omega_{(p)} = \norm{\omega_{(p)}}^2 \, \vol_{D3}~, \\
\norm{\omega_{(p)}}^2 \equiv \frac{1}{p!} \, (\omega_{(p)})_{\mu_1 \ldots \mu_p} (\omega_{(p)})_{\nu_1 \ldots \nu_p} \, g^{\mu_1 \nu_1} \cdots g^{\mu_p \nu_p}~,
\end{gather}
for the Hodge dual defined by the pullback metric $g$.

Thus, on our background, where $B_2, C_2, C_0$ all vanish, the D3 effective action takes the form
\begin{equation} \label{D3 action}
\cS_{D3} = -\tau_{D3} \int_\cW \Big( 1 + \bnorm{\cF}^2 + \frac14 \bnorm{\cF \wedge \cF}^2 \Big)^{1/2} \, \vol_{D3} + \tau_{D3} \int_\cW C_4.
\end{equation}
The worldvolume is now $\cW = \RR_t \times \widetilde \Delta_{ab}$.  Pulling back the metric~\eqref{IIB metric} to $\cW$, we obtain
\begin{equation} \label{D3 metric}
\begin{split}
\dd s_{D3} &= -\sqrt{Z_1 Z_2} \Big( (Z_1 Z_2 Z_3)^{-1} - Z_1^{-1} (\frv_1)^2 - Z_2^{-1} (\frv_2)^2 \Big) \dd t^2 - \frac{2 \mu}{Z_3 \sqrt{Z_1 Z_2}} \, \dd t \, \dd \psi \\
& \qquad + \frac{\cQ}{V^2 Z_3 \sqrt{Z_1 Z_2}} \, \dd \psi^2  + \frac{Z_3}{\sqrt{Z_1 Z_2}} \, (\dd v + A^3)^2 + V \sqrt{Z_1 Z_2} \, \dd z^2,
\end{split}
\end{equation}
and the determinant is
\begin{equation} \label{D3 det}
- \det g_{D3} = 1 - V^{-1} \cQ Z_1^{-1} (\frv_1)^2 - V^{-1} \cQ Z_2^{-1} (\frv_2)^2
~, \qquad 
\cQ \equiv Z_1 Z_2 Z_3 V - \mu^2 V^2~.
\end{equation}
The dynamics for $y_{5-8}$ are completely parallel to the M2-brane analysis, and lead to a Landau level structure for the brane motion on $\TT^4_{(5678)}$.  
The number of states in the lowest Landau level is thus given by $\Pi^{(1)}_{ab} \Pi^{(2)}_{ab}$, where $\Pi^{(I)}_{ab}$ is given by equation~\pref{Pi flux}.  
The $y_9, y_{10}$ dependence of the M2-brane dynamics are wrapped up into that of $\cF$; the details of how this works are given in Appendix~\ref{M2-D3 duality}, where we chase through the duality map.  For now, we content ourselves with counting the ground states available to the D3 worldvolume gauge field, to see that it matches the factor coming from the $\TT^2_{(9\,10)}$ dynamics of the M2-brane.  The counting depends on the topology of $\widetilde\Delta_{ab}$.

\subsubsection{The nature of the 3-cycles $\widetilde\Delta_{ab}$}

We want to evaluate the action of a probe D3-brane wrapping a three-cycle $\widetilde\Delta_{ab}$, stretching between charge centers $\yy_a$, $\yy_b$, in the D3-D3-P background given by \eqref{IIB metric}, \eqref{5form flux}.  The three-cycle $\widetilde\Delta_{ab}$ can locally be thought of as the two-cycle $\Delta_{ab}$ times the circle $\SS^1_v$ dual to $\TT^2_{(9\,10)}$, however $\SS^1_v$ is nontrivially fibered over the Gibbons-Hawking base.  The latter is itself a circle fibration of $\SS^1_\psi$ over $\RR^3$.  It will therefore be most useful to treat the two circles $\SS^1_\psi$ and $\SS^1_v$ on the same footing, as a $\TT^2$ fibration over $\RR^3$, and think of $\widetilde\Delta_{ab}$ as the total space of a path $\cC_{ab} \subset \RR^3$ between degeneration points of the fiber, together with the $\TT^2$ fiber.

Let us therefore re-organize the metric \eqref{IIB metric} to make such three-cycles more obvious.  Choosing the base $\cB_4$ to be a Gibbons-Hawking space with metric~\pref{Gibbons-Hawkingmetric}, with the functions $V, K^I, L_I, M$, as in~\pref{Vfn}-\pref{KLdefs} and harmonic one-forms
\begin{equation}
\beta \equiv B^3 \equiv \frac{K^3}{V} \, (\dd \psi + A) + \xi~, \qquad k = \mu \, (\dd \psi + A) + \omega~,
\end{equation}
one can then re-organize the 6d part of the metric (i.e. the first line of \eqref{IIB metric}) to look like an $\RR_t \times \TT^2$ fiber over $\RR^3$:
\begin{equation}
\label{6d metric explicit}
\begin{split}
\dd s_6^2 &= - \frac{2}{V \sqrt{Z_1 Z_2}} (\dd t + \omega) \Big[ V \, (\dd v + \xi) + K^3 \, (\dd \psi + A) \Big] \\
& \qquad + \frac{1}{V \sqrt{Z_1 Z_2}} \bigg[ \Big( -2 K^3 M + L_1 L_2 \Big) (\dd \psi + A)^2 + \Big( V L_3 + K^1 K^2 \Big) (\dd v + \xi)^2 \\
& \qquad \qquad \qquad \qquad + \Big( -2 V M - K^1 L_1 - K^2 L_2 + K^3 L_3 \Big) (\dd \psi + A) (\dd v + \xi) \bigg] \\
& \qquad + V \sqrt{Z_1 Z_2} \, \Big( \dd x^2 + \dd y^2 + \dd z^2 \Big)~.
\end{split}
\end{equation}
%

There are two basic ways in which the $\TT^2$ fiber can smoothly degenerate.  One is when the function $V$ blows up, while the $Z_I$ remain finite.  Then the $\psi$ circle pinches off smoothly.  This is given by
\begin{equation} \label{psi shrink}
V \sim \frac{q}{r}~, \qquad K^1 \sim \frac{k^1}{r}~, \qquad K^2 \sim \frac{k^2}{r}~, \qquad L_3 \sim -\frac{k^1 k^2}{q r}~,
\end{equation}
while the rest of the functions remain finite.  Another smooth degeneration is given by $Z_1, Z_2$ blowing up, while $V$ and $Z_3$ remain finite.  
Then the $v$ fiber pinches off smoothly.  This is described by
\begin{equation} \label{v shrink}
K^3 \sim \frac{k^3}{r}~, \qquad L_1 \sim \frac{\ell_1}{r}~, \qquad L_2 \sim \frac{\ell_2}{r}~, \qquad M \sim \frac12 \frac{\ell_1 \ell_2}{k^3 r}~,
\end{equation}
while the rest of the functions remain finite.%
\footnote{The harmonic function choices~\pref{KLdefs} become ill-defined when they have poles at locations where $V$ is regular; this ambiguity is resolved in Appendix~\ref{6d geom details}, where the appropriate generalization of the values of the fluxes $\Pi^{(I)}_{ab}$ are also given.} 
By combining the conditions \eqref{psi shrink} and \eqref{v shrink}, one can also arrange that some \emph{linear combination} of the $\psi$ and $v$ cycles smoothly pinches off.

Therefore, smooth solutions are determined by placing a collection of points $\yy_a$ in $\RR^3$ and, at each point, imposing some combination of \eqref{psi shrink} and \eqref{v shrink} such that the $m_a \psi + n_a v$ cycle pinches off smoothly there (modulo orbifold singularities, of course), for some integers $(m_a, n_a)$.  
As we will show below,
the topology of the cycle $\widetilde\Delta_{ab}$ is indicated by the determinant
\begin{equation}
\label{lens order}
\nu_{ab} = 
\begin{vmatrix}q_a & k^3_a \\ q_b & k^3_b \end{vmatrix} = q_a q_b \Pi^{(3)}_{ab} = 
\begin{cases} 0 ~~ , & \SS^2 \times \SS^1~; \\ p ~~ , & \SS^3/\ZZ_p~~~. \end{cases}
\end{equation}
The duality to type IIB has converted the four-form dipole flux $\cG_4$ along $\Delta_{ab}\times \TT^2_{(9\,10)}$ in the M-theory frame into the geometry of the $\TT^2$ fibration of the $\psi$ and $v$ circles over~$\RR^3$.  This relation is a variant of the duality between KK monopoles and NS5-branes%
~\cite{Gregory:1997te}.

The cycle $\widetilde\Delta_{ab}$ thus generically has the topology of a lens space $\SS^3/\ZZ_{p}$.
States of a $U(1)$ gauge field $\Da$ on such a space lie in topological sectors characterized by the first Chern class, which takes values in 
\be
H^2(\SS^3/\ZZ_p,\ZZ) = \ZZ_p ~.
\ee
Ground states are flat connections, specified by a choice of monodromy 
\be
\exp\Big[ i\int_\Sigma {\mathfrak \Da} \Big] = e^{2\pi i k/p}
\quad,\qquad k\in \{0,\ldots,p-1\}
\ee
around the fundamental torsion cycle $\Sigma$.  
Thus there are $p$ distinct ground states of the gauge field.%
\footnote{It seems that two phenomena are dual to one another: (1) the inability to simultaneously measure the momenta of the M2-brane along the two directions of the two-torus $\TT^2_{(9\,10)}$ in the M-theory frame; and (2) the inability%
~\cite{Gukov:1998kn,Freed:2006ya,Freed:2006yc}
to simultaneously measure the electric and magnetic fluxes of the D3-brane gauge field in the type IIB frame.}

Combining the number $p=|\nu_{ab}|$ of ground states of the gauge field, equation~\pref{lens order}, with the number of states $\Pi^{(1)}_{ab} \Pi^{(2)}_{ab}$ in the lowest Landau level of motion along $\TT^4$, one reproduces the M-theory frame result~\pref{hypernumber} for the overall number of ground states of the wrapped D3-brane.

\smallbreak

It remains to demonstrate that the topology of $\widetilde\Delta_{ab}$ is indeed that of a lens space of the appropriate order.  First, let us rewrite the metric~\eqref{6d metric explicit} in a way that makes the symmetries of the $\TT^2$ fibration more manifest.  This metric is invariant under the {\it spectral interchange operation} which exchanges the $\psi$ and $v$ circles, and therefore exchanges the harmonic functions
\begin{equation}
\label{spectral interchange}
V \leftrightarrow K^3, \qquad K^1 \leftrightarrow L_2, \qquad K^2 \leftrightarrow L_1, \qquad L_3 \leftrightarrow -2 M~.
\end{equation}
To make this symmetry manifest, define the spectral-interchange-invariant quantities
\begin{gather}
\hat H \equiv \sqrt{(V Z_1) (V Z_2)}\quad, \qquad  \cQ \equiv Z_1 Z_2 Z_3 V - \mu^2 V^2~, \\
\alpha \equiv V \, (\dd v + \xi) + K^3 \, ( \dd \psi + A)~, \\
\gamma \equiv (K^3)^2 \tilde \mu \, (\dd \psi + A) - V^2 \mu \, (\dd v + \xi)~.
\end{gather}
where in turn the functions $\mu$ and $\tilde \mu$, dual to each other under spectral interchange, are given by
\begin{align}
\mu &= M + \frac{1}{2 V} K^I L_I + \frac{K^1 K^2 K^3}{V^2}~, \\
\tilde \mu &= -\frac12 L_3 + \frac{1}{2 K^3} \big( K^1 L_1 + K^2 L_2 - 2V M\big) + \frac{L_1 L_2 V}{(K^3)^2}~.
\end{align}
The metric \eqref{6d metric explicit} is then given in terms of these quantities as%
~\cite{Niehoff:2013kia}
\begin{equation} \label{6d metric warner}
\dd s_6^2 = - 2 \hat H^{-1} \, (\dd t + \omega) \, \alpha + \hat H^{-3} \Big( \cQ \, \alpha^2 + \gamma^2 \Big) + \hat H \, \dd \vec y \cdot \dd \vec y
~.
\end{equation}

The advantage of rearranging the metric in the form \eqref{6d metric warner} is that it explicitly shows which cycle of the $T^2$ shrinks at each of the special points in the $\RR^3$ base.  Specifically, if we impose the following regularity conditions near a point $\yy_a$:
\begin{equation} \label{generic reg cond}
\hat H, \; \cQ, \; \alpha, \; \gamma \sim \frac{1}{|\yy-\yy_a|} \qquad \text{as} \quad \yy \to \yy_a~,
\end{equation}
then one has a point where the $U(1)$ fiber described by $\gamma$ pinches off smoothly.%
\footnote{It appears possible to allow double poles which cancel between $\cQ \alpha^2$ and $\gamma^2$; however, if the bubble equations are satisfied, then \eqref{generic reg cond} is the only possible choice.  One can show using the form of the metric \eqref{6d metric explicit} that the bubble equations are necessary for both causality and regularity, as demonstrated in Appendix \ref{bubble eqns proof}.}
Depending on the choices of parameters in the various functions $V, K^I, L_I, M$, this can be either a $v$ cycle, or a $\psi$ cycle, or some linear combination.  The metric has been arranged such that the linear combination $\gamma$ always describes the degenerating cycle, while $\alpha$ is non-degenerate.

Expanding near a pole location $\yy_a$, at leading order 
\begin{equation}
\alpha \sim {\it const.}\times \frac{(k^3_a \, \dd \psi + q_a \, \dd v)}{|\yy-\yy_a|} ~.
\end{equation}
Consider now the three-cycle $\widetilde\Delta_{ab}$ formed by $\TT^2_{(v,\psi)}$ fibered over a path $\cC_{ab}$ between $\yy_a$ and $\yy_b$.  The coordinate identifications of $(v,\psi)\in \RR^2$ can be made in two stages. Sitting at the pole $\yy_a$ and walking along the direction $\alpha$, we come back to the same point after a coordinate displacement $4\pi(q_a,k^3_a)$; similarly at the other pole.  There is thus a lattice of identifications $\Lambda$ of $(v,\psi)\in \RR^2$ 
\be
(v,\psi) \sim (v,\psi) + 4\pi m(q_a,k^3_a) + 4\pi n(q_b,k^3_b)
\quad,\qquad m,n\in\ZZ
\ee
If we start with $\cC_{ab}\times (\RR^2/\Lambda)$, and note that the cycles corresponding to $\gamma_{a,b}$ degenerate at the two poles, we get a submanifold $\widehat\Delta_{ab}$ that is topologically $\SS^3$ embedded in a 6d spacetime with metric~\eqref{6d metric explicit} where the periodicities of $(v,\psi)$ are given by $\Lambda$.  

However, the microstate geometry with metric~\eqref{6d metric explicit} has coordinate periodicities $(\delta v,\delta\psi)=4\pi (m,n)$ defining a lattice $\Lambda_0$, and this results in further identifications on $\widehat\Delta_{ab}$ that turn it into a lens space $\widetilde\Delta_{ab}$.%
\footnote{When $\gcd(q,k^3)=r>1$, the ambient 6d spacetime has a $\ZZ_r$ orbifold singularity at the corresponding pole~\cite{Giusto:2012yz}.}  
Specifically, one has
\be
\Lambda\approx |\nu_{ab}| \,\Lambda_0
\quad,\qquad
\nu_{ab} = q_b k^3_a - q_a k^3_b
~.
\ee
Closed paths on $\widetilde\Delta_{ab}$ corresponding to the elements of $\Lambda_0$ modulo $\Lambda$ are torsion cycles of order at most $\nu_{ab}$.  One can choose a representative torsion cycle of order $p=|\nu_{ab}|$ that generates all the others, and so $\widetilde\Delta_{ab}\approx \SS^3/\ZZ_p$ as advertised.


\section{Discussion}
\label{sec:Discussion}

Microstate geometries bear a remarkable similarity in their structure to little string theory on its Coulomb branch, in large part because the two are different scaling limits of BPS configurations of string theory on $\TT^5$.  This similarity suggests a close connection between the little string that accounts for the entropy of near-extremal NS5-branes, and the long string that accounts for the entropy of BTZ black holes in $AdS_3\times \SS^3$%
~\cite{Martinec:2014gka}.
The analogy furthermore suggests that the $AdS_3$ long string will make its appearance as the dominant excitation at the degeneration points of the Coulomb branch where charge centers coalesce, and that to find it one should consider the ``W-branes'' that stretch between charge centers and become light in the degeneration limit.

The W-branes in microstate geometries wrap cycles which degenerate when charge centers coalesce, and the geometry develops a throat of asymptotically large redshift.  BPS configurations of second-quantized W-branes have an entropy which scales with the same power of the charges as the black hole or black ring entropy, and accounts for a non-negligible fraction of that entropy.  The quiver quantum mechanics model used to obtain these results kept only the motion of the charge centers and the minimally wrapped W-branes in the effective dynamics.  Among the degrees of freedom omitted from this effective dynamics were those that would allow background charges and fluxes to rearrange themselves%
~\cite{Polchinski:1997pz}, and erase the information about the history of how the black hole was assembled from separated primitive bound states.  One might expect to recover more and more of the entropy as these and other degrees of freedom are introduced into the effective theory, provided this can be done in a controlled manner.

The attempt to assemble a BPS black hole in this fashion has been argued to lead to a singularity at the (inner) horizon%
~\cite{Marolf:2010nd,Marolf:2011dj,Murata:2013daa}.  One might therefore regard the appearance of all these light degrees of freedom as a rather sophisticated illustration of the basic principle of singularity resolution in string theory -- that UV singularities are resolved by IR physics of new light objects in the spectrum, which appear along with the would-be singularity and render it harmless.  
Indeed, these light degrees of freedom are typically branes wrapping a vanishing cycle%
~\cite{Strominger:1995cz,Hull:1995mz}.
The differences here are that there is a macroscopic number of light brane states associated to the relevant cycles, and that the proper spatial volume of the cycle does not vanish; instead, it becomes null from the perspective of an observer at spatial infinity.

One of the aspects of these results we find most intriguing is the idea that the long string of the CFT description of black hole states is a tangible object on the gravity side of the duality, and this notion raises the hope that a better understanding of its dynamics may help to resolve some of the long-standing puzzles of black hole dynamics (see%
~\cite{Martinec:2014gka} for a set of proposals).

Where might the long string be hiding in this picture?  It should be found in the configurations that give rise to the entropy.  The BPS degeneracy of pure-Higgs states of the triangular quiver maps to the counting of {\it 3-derangements}%
~\cite{Bena:2012hf}.
A $q$-derangement is a permutation of a set of $n$ copies of $q$ types of object such that no element of the set is mapped to an object of the same type.  Specifically, a 3-derangement is a permutation of the set 
\be
(\;\underbrace{1,\dots,1}_{\text{$a$~copies}}, \underbrace{2,\dots,2}_{\text{$b$~copies}}, \underbrace{3,\dots,3}_{\text{$c$~copies}}\;)
\ee
with $a+b+c=n$, such that no element of type $i\in\{1,2,3\}$, is mapped to another element of type $i$.  Such 3-derangements are equivalent to collections of closed paths on the triangular quiver, where each step along the path traverses a link on the quiver from one node to a different node.  At the same time, the quiver links are associated to wrapped W-branes; one may think of these paths as a collection of ``sausage-links'' made out of wrapped branes, joined at the poles $\yy_a$ where one cycle intersects another.  Could such bound states be the emergent configurations of the long string?

Finally, let us return to a lingering question from the Introduction: Are microstate geometries actually black holes?  Arguments have been made both in favor of and against this notion, see for instance%
~\cite{Bena:2013dka} and%
~\cite{Sen:2009bm,Banerjee:2009uk}.
There are several remarks to be made in light of our results:
\begin{enumerate}
\item
The analogy with NS5-branes on the Coulomb branch suggests that microstate geometries are not quite black holes, but they are on the cusp of becoming one.  On the Coulomb branch of NS5-branes, the little string is heavy, and the entropy consists of Hagedorn states of perturbative strings.  The entropy of these states does not scale with the charges as rapidly as the black fivebrane entropy, which is larger by a factor of order $\sqrt{n_5}$ due to the disparity between the tensions of the fundamental string and the little string.  Similarly, the entropy of the largest known class of horizonless microstate geometries%
~\cite{Bena:2010gg} 
is based on placing supertubes at the bottom of a microstate geometry with a long but finite throat and absorbing them into the background; the result does not scale with the charges as rapidly as the BTZ black hole entropy (though that deficiency could be due to an insufficiently general ansatz for the shape modes of the supertube%
~\cite{Niehoff:2012wu,Bena:2014qxa,Bena:2015bea}).
At the same time, the long string is not apparent.
\item
On the other hand, in the quiver quantum mechanics describing multicenter bound states, there is a Higgs-Coulomb equivalence for some states%
~\cite{Denef:2002ru,Sen:2011aa,Bena:2012hf}
Perhaps there might be a geometrization of the pure-Higgs states as well, once one takes into account the backreaction of the W-branes wrapping the vanishing cycles.  In such a situation, there would be dual interpretations of all the microstates as both geometries and brane configurations.  Alternatively, some fixed fraction of the degrees of freedom carrying the entropy might be geometric%
\footnote{For instance, certain R-symmetry current collective modes of the dual CFT fermions have been identified in a class of examples%
~\cite{Giusto:2012yz,Bena:2014qxa,Bena:2015bea,Giusto:2015dfa}
as shape modes in microstate geometries.}
and others not; the generic bound state would be a combination of the two.
\item
At the moment, the entropy of enumerated microstate geometries does not scale with the charges as rapidly as the three-charge black hole entropy.  Nevertheless, it is a possibility that techniques to construct a finite fraction of three-charge black hole states might be found, so that similar amounts come from geometrical structures and from horizon-wrapping brane condensates.  
Indeed, a proposal for the ``missing'' entropy of geometrical solutions has been explored in%
~\cite{Niehoff:2012wu,Bena:2014qxa,Bena:2015bea}.
But even if microstate geometries turn out not to be the main contributors to the density of states, the macroscopic entropy already found from geometrical constructions%
~\cite{Bena:2010gg}, 
the matching of certain low-energy black hole emission spectra%
~\cite{Chowdhury:2007jx,Chowdhury:2008uj,Chakrabarty:2015foa}, 
the possibility of perturbing them slightly to become black holes, and the increasingly precise duality map 
(see for example%
~\cite{Kanitscheider:2006zf,Giusto:2012yz,Bena:2015bea,Giusto:2015dfa})
make these objects particularly worthy of further study.
\end{enumerate}

In the quiver truncation, the dynamics of the collective modes of the charge centers and the lightest branes that stretch between them is well-behaved in the scaling limit.  Rather than leading to a continuous spectrum that one might naively expect from a scale invariant throat, the Higgs branch of the wrapped brane condensate has a finite volume configuration space and thus a gapped spectrum.  It is of considerable interest to understand how, in the full theory, the back-reacted condensate interacts with the supergravity modes to gap all the throat excitations, since naively there are arbitrarily many soft excitations of the geometry of an infinitely deep throat.  Similarly, the truncation to the minimally wrapped branes has left out all the shape excitations of the wrapped branes, and it is essential that these not also lead to an overcounting of the phase space available to extremal and near-extremal black holes.  We hope to address these issues in future work.


\section*{Acknowledgements}

We would like to thank Iosif Bena, Nikolay Bobev, Gary Gibbons, Sungjay Lee, Samir Mathur, David Turton, Nicholas Warner and Kenny Wong for useful discussions; 
and the Institut de Physique Theorique, Saclay, and the Centro de Ciencias de Benasque Pedro Pascual, for hospitality during the course of this work.
The work of EJM is supported in part by DOE grant DE-SC0009924.  The work of BEN is supported by ERC grant ERC-2011-StG 279363-HiDGR.



\appendix

\section{T-duality and reduction conventions}
\label{T duality conventions}

Throughout this work we make frequent changes of duality frame.  One can find reduction ans\"atze and expressions of the Buscher rules in many sources (e.g. \cite{Giveon:1994fu,Becker:2007zj,Johnson:2003gi}), although varying conventions are used, and formulas are typically given in index notation.  Here we find it convenient to use form notation, so it is useful to collect all of our conventions in one place, in form notation for further reference.

\subsection{Background fields: M-theory to IIA}

The reduction ansatz from M-theory to IIA is simple.  Write the 11d metric and 3-form potential as
\begin{equation}
\begin{split} \label{reduction ansatz}
\dd s_{11}^2 &= e^{-2\phi/3} \, \dd s_\text{IIA}^2 + e^{4\phi/3} \, (\dd y_{10} - C_1)^2, \\
\cA_3 &= C_3 + B_2 \wedge \dd y_{10},
\end{split}
\end{equation}
for a function $\phi$ on $\dd s_\text{IIA}^2$, and a 1-form $C_1$, 2-form $B_2$, and 3-form $C_3$, with legs along $\dd s_\text{IIA}^2$.  Then $B_2$ becomes the Kalb-Ramond field, and $C_1, C_3$ become the RR 1-form and 3-form potentials.  The sign choices are made such that the field strength has the reduction ansatz
\begin{equation}
\cF_4 \equiv \dd \cA_3 = G_4 + H_3 \wedge (\dd y_{10} - C_1), \qquad G_4 \equiv \dd C_3 + H_3 \wedge C_1,
\end{equation}
and the resulting 10d supergravity matches the conventions of \cite{Becker:2007zj}.

In our specific case, it is clear that $C_1 = 0$, and the remaining fields are given by
\begin{align}
\begin{split} \label {IIA metric}
\dd s_\text{IIA}^2 &= - \frac{1}{Z_3 \sqrt{Z_1 Z_2}} \, (\dd t + k)^2 + \sqrt{Z_1 Z_2} \, \dd s_4^2 \\
& \qquad \qquad \qquad + \sqrt{\frac{Z_2}{Z_1}} \big( \dd y_5^2 + \dd y_6^2 \big) + \sqrt{\frac{Z_1}{Z_2}} \big( \dd y_7^2 + \dd y_8^2 \big)  + \frac{\sqrt{Z_1 Z_2}}{Z_3} \, \dd y_9^2,
\end{split} \\
e^\phi &= \bigg( \frac{Z_1 Z_2}{Z_3^2} \bigg)^{1/4}, \qquad B_2 = A^3 \wedge \dd y_9, \qquad C_3 = A^1 \wedge \dd y_5 \wedge \dd y_6 + A^2 \wedge \dd y_7 \wedge \dd y_8. \label{IIA fields}
\end{align}

\subsection{Buscher rules}

To T-dualize from IIA to IIB (or vice versa) along a coordinate $y_9$, the Buscher rules for the NS sector fields can be summarized as
\begin{align}
\label{T-duality rules}
\dd s_{IIA}^2 &= \dd s_9^2 + f \, (\dd y_9 + \theta)^2~, 
& & & \dd \tilde s_{IIB}^2 &= \dd s_9^2 + \frac{1}{f} \, (\dd \tilde y_9 + B_2^1)^2~,   
\nonumber\\
B_2 &= B_2^2 + B_2^1 \wedge (\dd y_9 + \theta)~, 
& &\Longrightarrow & \tilde B_2 &= B_2^2 - \theta \wedge B_2^1 + \theta \wedge (\dd \tilde y_9 + B_2^1)~, 
\nonumber\\
e^\phi~, & & & & e^{\tilde \phi} &= f^{-1/2} e^\phi~.
\end{align}
Here $f$ is a function on $\dd s_9^2$, and $\theta$ is a 1-form, $B_2^1$ is a 1-form, and $B_2^2$ is a 2-form, all with legs on $\dd s_9^2$ only.  The dual coordinate in the IIB frame is $\tilde y_9$.  In our particular case, the 1-form $\theta = 0$, and we can write the NS sector fields as
\begin{align}
\begin{split} \label{IIB metric A}
\dd s_\text{IIB}^2 &= - \frac{1}{Z_3 \sqrt{Z_1 Z_2}} \, (\dd t + k)^2 + \sqrt{Z_1 Z_2} \, \dd s_4^2 + \frac{Z_3}{\sqrt{Z_1 Z_2}} \big( \dd \tilde y_9 +A^3 \big)^2 \\
& \qquad \qquad \qquad + \sqrt{\frac{Z_2}{Z_1}} \big( \dd y_5^2 + \dd y_6^2 \big) + \sqrt{\frac{Z_1}{Z_2}} \big( \dd y_7^2 + \dd y_8^2 \big),
\end{split} \\
e^\phi &= 1, \\
\tilde B_2 &= 0.
\end{align}

For the RR fields, it is most straightforward to T-dualize the field strengths, and then work backwards to find the corresponding potentials (which will be needed in the DBI and WZ actions).  We define the field strengths $G_{p+1}$ via
\begin{equation}
G_{p+1} \equiv \dd C_p + H_3 \wedge C_{p-2}.
\end{equation}
It is then simplest to break up the field strengths in a way similar to what we did above for the B field.  The IIA field strengths are
\begin{equation}
G_2 \equiv G_2^2 + G_2^1 \wedge (\dd y_9 + \theta), \qquad G_4 \equiv G_4^4 + G_4^3 \wedge (\dd y_9 + \theta),
\end{equation}
where the $G$'s are forms on $\dd s_9^2$; the lower index indicates to which $p$-form field strength each $G$ belongs, and the upper index indicates the degree of each form.  Likewise, the IIB RR fields are
\begin{gather}
G_1 \equiv G_1^1 + G_1^0 \, (\dd \tilde y_9 + B_2^1), \qquad G_3 \equiv G_3^3 + G_3^2 \wedge (\dd \tilde y_9 + B_2^1), \\
G_5 \equiv G_5^5 + G_5^4 \wedge (\dd \tilde y_9 + B_2^1).
\end{gather}
When the forms are written in this way, the T-duality rules simply set each pair of $G$'s equal to each other, which have the same degree.  That is, the Buscher rules for the RR field strengths become\footnote{We note here that $G_1^0 = G_0$ in general; that is, it is the T-dual of the Romans mass.}
\begin{equation}
G_1^0 = 0, \qquad G_1^1 = G_2^1, \qquad G_3^2 = G_2^2, \qquad G_3^3 = G_4^3, \qquad G_5^4 = G_4^4,
\end{equation}
and the $G_5^5$ part of $G_5$ is fixed by self-duality:
\begin{equation}
G_5^5 = \thodge_{10} \Big( G_5^4 \wedge (\dd \tilde y_9 + B_2^1) \Big).
\end{equation}
Applying these rules in our specific case, we obtain $G_1 = 0$, $G_3 = 0$, and
\begin{equation}
G_5 = (1 + \thodge_{10}) \bigg[ \Big( \dd A^1 \wedge \dd y_5 \wedge \dd y_6 + \dd A^2 \wedge \dd y_7 \wedge \dd y_8 \Big) \wedge \big( \dd v + A^3 \big) \bigg].
\end{equation}
This can be written out as
\begin{equation} \label{5form flux A}
\begin{split}
G_5 &= \Big( \dd A^1 \wedge \dd y_5 \wedge \dd y_6 + \dd A^2 \wedge \dd y_7 \wedge \dd y_8 \Big) \wedge \big( \dd v + A^3 \big) \\
& \qquad - \Big[ Z_1^{-1} Z_3^{-1} \, (\dd u + k) \wedge \big( Z_2 \hodge_4 \Theta^2 - \hodge_4 \dd k \big) + \hodge_4 \dd Z_2 \Big] \wedge \dd y_5 \wedge \dd y_6 \\
& \qquad - \Big[ Z_2^{-1} Z_3^{-1} \, (\dd u + k) \wedge \big( Z_1 \hodge_4 \Theta^1 - \hodge_4 \dd k \big) + \hodge_4 \dd Z_1 \Big] \wedge \dd y_7 \wedge \dd y_8.
\end{split}
\end{equation}
Using the BPS equations \eqref{BPS eqns}, we verify that $\dd G_5 = 0$, and hence we can in principle integrate this to obtain a $C_4$ such that $G_5 = \dd C_4$.  There are many choices of gauge; we will find it convenient to choose a gauge such that the first line reads
\begin{equation}
C_4 = \Big( y_5 \, \dd A^1 \wedge \dd y_6 + y_7 \, \dd A^2 \wedge \dd y_8 \Big) \wedge \big( \dd v + A^3 \big) + \cdots.
\end{equation}

\subsection{Probe brane effective actions}

In order to chase the M2-brane degrees of freedom through the chain of dualities, we must work out the explicit effects of the above transformations on the M2 brane probe action.  We will simply include results in form language; we refer the reader to \cite{Simon:2011rw} for a derivation.

The M2 brane effective action is given by
\begin{equation} \label{M2 action A}
\cS_{M2} = - \tau_{M2} \int \vol_{M2} + \tau_{M2} \int \cA_3.
\end{equation}
Choosing coordinates $\xi^i$ on the M2, one can write the action in the more usual way:
\begin{equation}
\cS_{M2} = -\tau_{M2} \int \dd^3 \xi \, \sqrt{- g_{M2} } + \tau_{M2} \int \cA_3,
\end{equation}
where $g_{M2} \equiv i^* g_{11}$ for the pullback along the embedding $i : M2 \hookrightarrow \cM_{11}$.

Next we apply the reduction to IIA.  When the M2-brane is not extended along $y_{10}$ (as in our case), one obtains a ``direct dimensional reduction" to the D2 action
\begin{equation}
\cS_{D2} = - \tau_{D2} \int e^{-\phi} \Big( 1 + \bnorm{B_2 + \cF}_{D2}^2 \Big)^{1/2} \, \vol_{D2} + \tau_{D2} \int e^{B_2 + \cF} \wedge \cC,
\end{equation}
which can alternatively be written
\begin{equation}
\cS_{D2} = - \tau_{D2} \int \dd^3 \xi \, e^{-\phi} \sqrt{ - \det (g_{10} + B_2 + \cF) } + \tau_{D2} \int e^{B_2 + \cF} \wedge \cC,
\end{equation}
where the usual pullbacks to the D2 are understood.  The D2 tension $\tau_{D2} = \tau_{M2}$.  The original $y_{10}$ degree of freedom has been exchanged for a worldvolume gauge field with field strength $\cF \equiv 2 \pi \alpha' \, F$, which is related to $y_{10}$ via a nonlinear duality:
\begin{equation} \label{y10 in F}
\dd y_{10} - C_1 = e^{-\phi} \, \Big( 1 + \bnorm{B_2 + \cF}_{D2}^2 \Big)^{-1/2} \hodge_{D2} \big(B_2 + \cF).
\end{equation}
In this formula, we treat $y_{10}$ as a scalar field living on the D2-brane.

To move from IIA to IIB, we apply the T-duality rules.  Again, the D2 brane is not extended along the duality coordinate $y_9$, so it gains a leg along the new coordinate $\tilde y_9$.  The probe effective action becomes
\begin{equation} 
\cS_{D3} = -\tau_{D3} \int e^{- \tilde \phi} \Big( 1 + \bnorm{\tilde b}^2 + \frac14 \bnorm{\tilde b \wedge \tilde b}^2 \Big)^{1/2} \, \vol_\text{D3} + \tau_{D3} \int e^{\tilde b} \wedge \cC,
\end{equation}
where $\tilde b \equiv \tilde B_2 + \tilde \cF$, which can equivalently be written
\begin{equation}
\cS_{D3} = - \tau_{D3} \int \dd^4 \xi \, e^{-\tilde \phi} \sqrt{ - \det (\tilde g_{10} + \tilde B_2 + \tilde \cF) } + \tau_{D3} \int e^{\tilde B_2 + \tilde \cF} \wedge \cC.
\end{equation}
The D-brane tensions $\tau_{D2}, \tau_{D3}$ are related in the usual way\footnote{Restoring appropriate units, one must obtain $R = \sqrt{\alpha'}$.},
\begin{equation}
\tau_{D2} = \tau_{D3} \oint \dd \tilde y_9 = 2 \pi R \, \tau_{D3},
\end{equation}
and the gauge field strength $\cF = \dd \Da$ acquires a new term proportional to the velocity in the original $y_9$ direction:
\begin{equation}
\tilde \cF = \cF + \dd y_9 \wedge \dd \tilde y_9, \qquad \text{or} \qquad \tilde \Da = \Da + y_9 \, \dd \tilde y_9.
\end{equation}

\section{Details of the 6d geometry}
\label{6d geom details}

In order to write down specific SUGRA solutions, we must work out the consequences of the regularity conditions \eqref{generic reg cond} on the various parameters $q_a, k^I_a, \ell_I^a, m_a$ of the solution.  In principle, one can work this out directly, but using the spectral interchange symmetry~\eqref{spectral interchange}, there is a more efficient route.

We know that every point $r_a$ where $q_a \neq 0$, the regularity conditions must reduce to the usual 5d regularity conditions as given in \cite{Bena:2007kg}:
\begin{equation} \label{psi reg cond}
\ell_1^a = - \frac{k^2_a k^3_a}{q_a}, \qquad \ell_2^a = - \frac{k^1_a k^3_a}{q_a}, \qquad \ell_3^a = - \frac{k^1_a k^2_a}{q_a}, \qquad m_a = \frac{k^1 k^2 k^3}{2 q_a^2},
\end{equation}
together with the bubble equations
\begin{gather}
- 2 q_a m_0 + 2 \varepsilon_0 m_a - k^I_a \ell_I^0 + k^I_0 \ell_I^a = \sum_{b \neq a} \frac{q_a q_b}{|\yy_a-\yy_b|} \Pi^{(1)}_{ab} \Pi^{(2)}_{ab} \Pi^{(3)}_{ab}, \\
\Pi^{(I)}_{ab} \equiv \frac{k^I_b}{q_b} - \frac{k^I_a}{q_a}
~.
\end{gather}
These conditions were implicitly assumed in the choices~\eqref{KLdefs} for the harmonic fuctions.
While in the 5d geometry of the M-theory frame, the bubble equations came from demanding the local absence of closed timelike curves near the $r_a$, in the 6d geometry of the IIB frame they come from imposing the regularity conditions~\eqref{generic reg cond}, particularly on the one-form $\gamma$.

In the case where both $q_a \neq 0$ and $k^3_a \neq 0$, the conditions \eqref{psi reg cond} can be algebraically rearranged:
\begin{equation} \label{v reg cond}
k^1_a = - \frac{q_a \ell_2^a}{k^3_a}
~, \qquad 
k^2_a = -\frac{q_a \ell_1^a}{k^3_a}
~, \qquad 
m_a = \frac{\ell_1^a \ell_2^a}{2 k^3_a}
~, \qquad 
\ell_3^a = - \frac{q_a \ell_1^a \ell_2^a}{(k^3_a)^2}
~,
\end{equation}
which can be seen as the spectral-interchange dual to \eqref{psi reg cond}.  The difference is that \eqref{v reg cond} can be applied in the case $q_a = 0$.  The bubble equations can be transformed in a similar way, noting that for generic parameters
\begin{equation}
\frac{k^1_a}{q_a} = - \frac{\ell_2^a}{k^3_a}, \qquad \frac{k^2_a}{q_a} = - \frac{\ell_1^a}{k^3_a}, \qquad q_a q_b \Pi^{(3)}_{ab} = q_a k^3_b - q_b k^3_a;
\end{equation}
however, the new expressions are valid in the case that $q_a = 0$.  So we can write the more general form of the bubble equations as
\begin{equation} \label{bubble eqns}
- 2 q_a m_0 + 2 \varepsilon_0 m_a - k^I_a \ell_I^0 + k^I_0 \ell_I^a = \sum_{b \neq a} \frac{1}{|\yy_a-\yy_b|} \Pi^{(1)}_{ab} \Pi^{(2)}_{ab} \, (q_a k^3_b - q_b k^3_a),
\end{equation}
where
\begin{equation} \label{Pi cases}
\Pi^{(1,2)}_{ab} \equiv P^{(1,2)}_b - P^{(1,2)}_a
\quad, \qquad 
P^{(1,2)}_a \equiv \begin{cases} ~~\dfrac{k^{1,2}_a}{q_a}~, \quad q_a \neq 0~, \\[2ex] -\dfrac{\ell_{2,1}^a}{k^3_a}~, \quad k^3_a \neq 0~. \end{cases}
\end{equation}
Then for generic values of the parameters, we can always use some combination of \eqref{psi reg cond}, \eqref{v reg cond}, and \eqref{bubble eqns} to enforce regularity.  Finally, the sum of all the bubble equations results in the condition
\begin{equation} \label{zero parts}
-2 \, \Big( \sum_a q_a \Big) \, m_0 + 2 \, \Big( \sum_a m_a \Big) \, \varepsilon_0 - \Big( \sum_a k^I_a \Big) \, \ell_I^0 + \Big( \sum_a \ell_I^a \Big) \, k^I_0 = 0,
\end{equation}
which relates to the asymptotic behavior at $r \to \infty$.

\subsection{Behavior at infinity}

To control the asymptotics of the solution, we look at the behavior of each function at infinity.  To get \emph{flat} asymptotics, the condition is the same as the near-point limit, except taken at infinity instead:
\begin{equation}
\hat H, \; \cQ, \; \alpha, \; \gamma \sim \frac{1}{r} \qquad \text{as} \quad r \to \infty.
\end{equation}
For $AdS_3 \times S^3$ asymptotics, one needs instead
\begin{equation}
\hat H \sim \frac{1}{r^2}, \quad \cQ \, \alpha \sim \frac{1}{r^4}, \quad \gamma \sim \frac{1}{r^3}, \qquad \text{as} \quad r \to \infty.
\end{equation}
This condition is obtained by looking at the 6d metric \eqref{6d metric warner} and demanding that
\begin{equation}
\hat H \, \dd \vec y \cdot \dd \vec y \sim \frac{\dd r^2}{r^2} + \dd \Omega_2^2, \qquad \hat H^{-3} \, \gamma^2 \sim \cO(r^0),
\end{equation}
such that the $S^3$ part is made of the fiber $\gamma$ over $\dd \Omega_2^2$, together with the condition that
\begin{equation}
- 2 \hat H^{-1} \, (\dd t + \omega) \, \alpha, \qquad \hat H^{-3} \, \cQ \, \alpha^2
\end{equation}
have the same behavior in $r$ (so that they combine with $\dd r^2/r^2$ to form the $AdS_3$ part).

\subsection{Two-center example:  $AdS_3 \times S^3$}

If we choose a two-center example with asymptotics as in the previous section, we should get $AdS_3 \times S^3$.  Let's put charges at $z = \pm a$, so that
\begin{equation}
r_\pm \equiv \sqrt{\rho^2 + (z \mp a)^2}
\end{equation}
in cylindrical coordinates.  Now choose the parameters
\begin{align}
q_+ &= 1~, & 
k^3_+ &= 0~, & 
k^{1,2}_+ &= 1~, & 
\ell_{1,2}^+ &= 0~, \\ 
q_- &= 0~, & 
k^3_- &= 1~, &
k^{1,2}_- &= 0~, & 
\ell_{1,2}^- &= 1~.
\end{align}
According to the regularity conditions, we must then have
\begin{align}
\ell_3^+ &= -1~, & \ell_3^- &= 0~, & m_+ &= 0~, & m_- &= \frac12~.
\end{align}
Then the bubble equations become
\begin{equation}
- 2 m_0 = \frac{1}{2a} \Pi^{(1)}_{+-} \Pi^{(2)}_{+-}~, \qquad - \ell_3^0 = \frac{1}{2a} \Pi^{(1)}_{-+} \Pi^{(2)}_{-+}~,
\end{equation}
where each of the fluxes is given by
\begin{equation}
\Pi^{(1)}_{+-} = - \frac{\ell_2^-}{k^3_-} - \frac{k^1_+}{q_+} = -2~, \qquad \Pi^{(2)}_{+-} = - \frac{\ell_1^-}{k^3_-} - \frac{k^2_+}{q_+} = -2~.
\end{equation}
The bubble equations can be solved by
\begin{equation} \label{const a}
\ell_3^0 = \frac{2}{a}~, \qquad m_0 = - \frac{1}{a}~,
\end{equation}
although we will show that the value of $a$ is pure gauge, provided that the bubble equations are satisfied, and can be set to $1$ by rescaling the coordinate $t$.

The functions appearing in the metric are given by
\begin{gather}
V = \frac{1}{r_+}~, \qquad K^3 = \frac{1}{r_-}~, \qquad K^{1,2} = \frac{1}{r_+}~, \qquad L_{1,2} = \frac{1}{r_-}~, \\
M = -\frac12 \bigg( \frac{2}{a} - \frac{1}{r_-} \bigg)~, \qquad L_3 = \frac{2}{a} - \frac{1}{r_+}~,
\end{gather}
and the various one-forms are given by
\begin{equation}
A = \frac{z-a}{r_+} \, \dd \phi~, \quad \xi = - \frac{z+a}{r_-} \, \dd \phi~, \quad \omega = \frac{1}{a} \bigg(1 + \frac{z-a}{r_+} - \frac{z+a}{r_-} - \frac{\rho^2 + z^2 - a^2}{r_+ r_-} \bigg) \dd \phi~.
\end{equation}
The pure gauge part $(1/a) \, \dd \phi$ of $\omega$ is necessary to eliminate Dirac string singularities along the axis, which otherwise would allow small closed timelike curves.

Plugging all of this into the metric \eqref{6d metric explicit} yields, after a fair bit of algebra,
\begin{equation} \label{6d example cylindrical}
\begin{split}
\dd s_6^2 &= - \dd t \, \Big( r_+ \, \dd \psi + r_- \, \dd v - 2a \, \dd \phi \Big) + \frac{2}{r_+ r_-} \Big( \dd \rho^2 + \dd z^2 \Big) + 2 \, \dd \phi^2 \\
& \qquad  + \frac{1}{a} \Big[ r_+ \, \dd \psi^2 + r_- \, \dd v^2 + \big( r_+ + r_- - 2a) \, \dd \psi \, \dd v \\
& \qquad \qquad \qquad + \big( - r_+ + r_- - 2a \big) \, \dd \psi \, \dd \phi + \big(r_+ - r_- - 2a \big) \, \dd v \, \dd \phi \Big]~.
\end{split}
\end{equation}
Next we change to bipolar coordinates defined via
\begin{equation}
\rho = a \sinh \xi \sin \theta~, \qquad z = a \cosh \xi \cos \theta~, \qquad r_\pm = a (\cosh \xi \mp \cos \theta)~.
\end{equation}
Immediately plugging this into \eqref{6d example cylindrical} yields the metric
\begin{equation} \label{6d example bipolar}
\begin{split}
\dd s_6^2 &= - a \, \dd t \, \Big[ \cosh \xi (\dd \psi + \dd v) - \cos \theta (\dd \psi - \dd v) - 2 \, \dd \phi \Big] + 2 \, \Big( \dd \xi^2 + \dd \theta^2 + \dd \phi^2 \Big) \\
& \qquad + (\cosh \xi - \cos \theta) \, \dd \psi^2 + (\cosh \xi + \cos \theta) \, \dd v^2 + 2 \, (\cosh \xi - 1) \, \dd \psi \, \dd v \\
& \qquad + 2 \, (\cos \theta -1) \, \dd \psi \, \dd \phi - 2 \, (\cos \theta + 1) \, \dd v \, \dd \phi~.
\end{split}
\end{equation}
Now we see explicitly that $a$ is a coordinate artifact.  With some effort, \eqref{6d example bipolar} can be rearranged to give the metric of $AdS_3 \times S^3$:
\begin{equation} \label{6d example final}
\dd s_6^2 = 2 \, \Big[ \dd \xi^2 + \sinh^2 \xi \, \dd \eta^2 - \big( \dd \tau - \cosh \xi \, \dd \eta)^2 \Big] + 2 \, \Big[ \dd \theta^2 + \sin^2 \theta \, \dd \varphi^2 + \big( \dd \chi - \cos \theta \, \dd \varphi \big)^2 \Big],
\end{equation}
after changing coordinates as follows:
\begin{equation}
2 \tau = \psi + v - a \, t~, \qquad 2 \eta = \psi + v~, \qquad 2 \chi = \psi - v~, \qquad 2 \varphi = \psi + v - a \, t - 2 \phi~.
\end{equation}

Now consider the above metric \eqref{6d example final} on a constant-$t$ slice, and pulled back to the interval $-a < z < a$ at $\rho = 0$.  Then $\xi = 0$, and the $AdS_3$ part of \eqref{6d example final} vanishes.  The limit $\rho \to 0$ is a little tricky, but for simplicity one can approach the limit along a $\phi = (\text{const})$ plane.  Then the metric becomes a 3-sphere described by the coordinates $(\theta, \psi, v)$:
\begin{equation}
\dd s^2 \to 2 \, \bigg\{ \dd \theta^2 + \sin^2 \theta \, \bigg( \frac{\dd \psi + \dd v}{2} \bigg)^2 + \bigg[ \bigg( \frac{\dd \psi - \dd v}{2} \bigg) - \cos \theta \, \bigg( \frac{\dd \psi + \dd v}{2} \bigg) \bigg]^2 \bigg\}~.
\end{equation}
So we see that the $S^3$ bubble described by the $v$ and $\psi$ fibers over the interval $-a < z < a$ in the original coordinates is homologous to the $S^3$ factor of $AdS_3 \times S^3$, as we should expect.

\subsection{The effect of violating the bubble equations}
\label{bubble eqns proof}

In the 5d geometry of the M-theory frame, violating the bubble equations results in small closed timelike curves near the Gibbons-Hawking centers $\yy_a$.  In the 6d geometry of the IIB frame, violating the bubble equations \emph{also} leads to small closed timelike curves, although it is less obvious how to see this.

First we can reuse the previous example to demonstrate.  Let us set up everything as before, except that we will neglect to impose \eqref{const a}.  If we pull back the 6d metric~\eqref{6d example cylindrical} to the ``bubble'' described by $(v, \psi, z)$, we obtain
\begin{equation}
\dd s_3^2 = \frac{2}{r_+ r_-} \, \dd z^2 + \frac{1}{a} \Big[ r_+ \, \dd \psi^2 + r_- \, \dd v^2 + \big( r_+ + r_- - 2a) \, \dd \psi \, \dd v \Big]~.
\end{equation}
If the bubble equations were imposed, we would have $r_+ + r_- = 2a$, and hence the last term would vanish.  In fact, the presence of this term spoils regularity \emph{and} causality at each endpoint.

Take for example the limit $r_- \to 0$.  Then the torus part of the metric looks like
\begin{equation}
r_+ \, \dd \psi^2 + (r_+ - 2a) \, \dd \psi \, \dd v~,
\end{equation}
where $r_+$ is now some constant.  The determinant of the above expression is
\begin{equation} \label{bad det}
- \frac14 (r_+ - 2a)^2~.
\end{equation}
From \eqref{bad det}, we make two troubling observations:  First, it fails to be zero!  So there is no zero eigenvalue, and in fact no fiber is pinching off at $r_- \to 0$ at all.  Second, it is actually \emph{negative}, which means that the metric acquires a negative eigenvalue along one of the $\TT^2_{(v, \psi)}$ directions near $r_- \to 0$.  The metric is both badly singular \emph{and} has closed timelike curves.\footnote{Of course, in order for an eigenvalue to go negative, it must pass through zero; there is some $\SS^1 \subset \TT^2_{(v, \psi)}$ that pinches off somewhere else, and the 6d metric is singular there, too.}

We can also work out the more general case which shows that the bubble equations are a necessity in 6d.  It is easiest to use the form of the metric \eqref{6d metric explicit}.  Near the Gibbons-Hawking points, one of $V, K^3$ must go like $1/r$, and hence to compensate in the first term of \eqref{6d metric explicit} we must have
\begin{equation}
\hat H \equiv V \sqrt{Z_1 Z_2} \sim r^{-1}
\end{equation}
just as before.  In order to have one fiber that shrinks, while the other fiber does not, the terms in brackets on the second line of \eqref{6d metric explicit} be no more singular than $r^{-1}$:
\begin{gather}
\Big( -2 K^3 M + L_1 L_2 \Big) \sim \frac{1}{r}~, \qquad \Big( V L_3 + K^1 K^2 \Big) \sim \frac{1}{r}~, \\
\Big( -2 V M - K^1 L_1 - K^2 L_2 + K^3 L_3 \Big) \sim \frac{1}{r}~.
\end{gather}
There is a catch, though.  Of the $\TT^2$ directions spanned by $v, \psi$, only \emph{one} of them should go as $1/r$, because that is precisely the one that does \emph{not} shrink.  The fiber that shrinks is subleading.  Therefore, while the portion in brackets on the second line of \eqref{6d metric explicit} must vanish at order $r^{-2}$, at order $r^{-1}$ it must \emph{degenerate} such that it can be written as one square.

At order $r^{-2}$, we have
\begin{gather}
\Big( -2 K^3 M + L_1 L_2 \Big) \sim \frac{1}{r^2} \Big( \ell_1^a \ell_2^a - 2 k^3_a m_a \Big)~, \\
\Big( V L_3 + K^1 K^2 \Big) \sim \frac{1}{r^2} \Big( q_a \ell_3^a + k^1_a k^2_a \Big)~, \\
\Big( -2 V M - K^1 L_1 - K^2 L_2 + K^3 L_3 \Big) \sim \frac{1}{r^2} \Big( -2 q_a m_a + k^3_a \ell_3^a - k^1_a \ell_1^a - k^2_a \ell_2^a \Big)~.
\end{gather}
Together with the condition $\hat H \sim 1/r$, these imply all of the local regularity conditions in \eqref{psi reg cond} and \eqref{v reg cond}.

Next we look at the $r^{-1}$ order.  For simplicity, we assume that all of the constant parts vanish except $\ell_3^0, m_0$.  It is more or less straightforward to put them back in.  After some algebra, the entire bracketed term can be written
\begin{equation} \label{brackets 1}
\begin{split}
[ \cdots ] &\sim \frac{1}{r} \, \Big[ k^3_a \, (\dd \psi + A) + q_a \, (\dd v + \xi) \Big] \times \\
& \times \bigg[ (\dd \psi + A) \bigg( - 2 m_0 - \sum_b \frac{k^3_b}{r_{ab}} \Pi^1_{ab} \Pi^2_{ab} \bigg) + (\dd v + \xi) \bigg( \ell_3^0 - \sum_b \frac{q_b}{r_{ab}} \Pi^1_{ab} \Pi^2_{ab} \bigg) \bigg]~.
\end{split}
\end{equation}
Next, using the identity
\begin{equation}
ax + by = \frac12 (a + b)(x + y) + \frac12 (a -b)(x-y)~,
\end{equation}
we can re-write \eqref{brackets 1} to read
\begin{equation} \label{brackets 2}
\begin{split}
[ \cdots ] &\sim \frac{1}{r} \, \Big[ k^3_a \, (\dd \psi + A) + q_a \, (\dd v + \xi) \Big] \times \frac{1}{q_a k^3_a} \times \\
& \times  \bigg\{ \frac12 \Big[ k^3_a \, (\dd \psi + A) + q_a \, (\dd v + \xi) \Big] \bigg( -2 q_a m_0 + k^3_a \ell_3^0 - \sum_b \frac{1}{r_{ab}} \Pi^1_{ab} \Pi^2_{ab} (q_a k^3_b + q_b k^3_a) \bigg) \\
& \qquad + \frac12 \Big[ k^3_a \, (\dd \psi + A) - q_a \, (\dd v + \xi) \Big] \bigg( -2 q_a m_0 - k^3_a \ell_3^0 - \sum_b \frac{q_a q_b}{r_{ab}} \Pi^1_{ab} \Pi^2_{ab} \Pi^3_{ab} \bigg) \bigg\}~.
\end{split}
\end{equation}
Now, observe that the 1-form in the second line of \eqref{brackets 2} matches the one in the first line, so this must give us our one square at order $r^{-1}$.  The 1-form in the second line of \eqref{brackets 2} is linearly independent; however, we recognize its coefficient as precisely the bubble equation.  Therefore, the bubble equation must be satisfied in order for the fiber to pinch off.  One can also show, in a computation similar to what was done in our 2-center example, that the determinant of \eqref{brackets 2} is negative unless the bubble equations are satisfied.

Furthermore, we confirm that the fiber which does \emph{not} pinch off (i.e., the first line of \eqref{brackets 2}) is the one proportional to $\alpha$ in \eqref{6d metric warner}.


\section{Details of M2-brane/D3-brane duality}
\label{M2-D3 duality}

Here, we investigate the duality chain that converts M2-brane dynamics in the background~\eqref{11d metric},~\eqref{11d 3form} to D3-brane dynamics in the type IIB background~\eqref{IIB metric},~\eqref{5form flux}.

\subsection{Reducing M2 branes to D2 branes}

We begin with the probe brane action for D2 branes in the IIA background that arises upon reduction from M-theory.  The probe brane action is given by
\begin{equation} \label{D2 action}
\cS_{D2} = - \tau_{D2} \int e^{-\phi} \Big( 1 + \bnorm{B_2 + \cF}_{D2}^2 \Big)^{1/2} \, \vol_{D2} + \tau_{D2} \int e^{B_2 + \cF} \wedge \cC~,
\end{equation}
where $\cF$ is related to the M2 worldvolume field $y_{10}$ of the M-theory frame via
\begin{equation} \label{y10 in F 2}
\dd y_{10} - C_1 = e^{-\phi} \, \Big( 1 + \norm{B_2 + \cF}_{D2}^2 \Big)^{-1/2} \hodge_{D2} \big(B_2 + \cF)~.
\end{equation}
The IIA background after reducing from M theory is given by~\eqref{IIA metric},~\eqref{IIA fields}.

Again the first step is to compute the pullback metric.  Taking the $y_i$ to be functions of $t$ only, we obtain
\begin{equation} \label{D2 metric pullback}
\begin{split}
\dd s_{D2}^2 &= - \sqrt{Z_1 Z_2} \, \Big( (Z_1 Z_2 Z_3)^{-1} - Z_1^{-1} (\frv_1)^2 - Z_2^{-1} (\frv_2)^2 - Z_3^{-1} (\dot y_9^2) \Big) \, \dd t^2 \\
& \qquad \qquad - \frac{2 \mu}{Z_3 \sqrt{Z_1 Z_2}} \, \dd t \, \dd \psi + \frac{\cQ}{V^2 Z_3 \sqrt{Z_1 Z_2}} \, \dd \psi^2 + V \sqrt{Z_1 Z_2} \, \dd z^2~,
\end{split}
\end{equation}
whose determinant, including dilaton factor, is
\begin{equation}
e^{-\phi} \sqrt{- \det g_{D2}} = \Big(1 - V^{-1} \cQ Z_1^{-1} (\frv_1)^2 - V^{-1} \cQ Z_2^{-1} (\frv_2)^2 - V^{-1} \cQ Z_3^{-1} (\dot y_9^2) \Big)^{1/2}~.
\end{equation}

At first glance, one might be skeptical that the action \eqref{D2 action} should produce the same dynamics for the $y_i$ as the M2-brane action \eqref{M2 action}, although it must by duality.  All of the apparent discrepancies are resolved by using the relation \eqref{y10 in F 2} between the gauge field strength $\cF$ and the original coordinate $y_{10}$.%
\footnote{Although we stress that the relation \eqref{y10 in F 2} cannot be imposed until \emph{after} varying the action, because $y_{10}$ is no longer a fundamental field, while the gauge potential $\Da$ is.}
We want to know what this relation is explicitly.  Taking the square of \eqref{y10 in F 2} and using the fact that $\hodge_{D2}^2 = -1$, it is easy to show that
\begin{equation}
\Big( 1 + \norm{B_2 + \cF}_{D2}^2 \Big) = \Big( 1 + e^{2 \phi} \norm{\dd y_{10}}_{D2}^2 \Big)^{-1}~.
\end{equation}
One can then compute $\norm{\dd y_{10}}^2$ using (the inverse of) the pullback metric~\eqref{D2 metric pullback}, and thus show, after a bit of algebra, that
\begin{equation}
\Big( 1 + \norm{B_2 + \cF}_{D2}^2 \Big) = \frac{ \Big( 1 - V^{-1} \cQ Z_1^{-1} (\frv_1)^2 - V^{-1} \cQ Z_2^{-1} (\frv_2)^2 - V^{-1} \cQ Z_3^{-1} (\dot y_9^2) \Big) }{ \Big( 1 - V^{-1} \cQ \sum_I Z_I^{-1} (\frv_I)^2 \Big) }~,
\end{equation}
and the worldvolume field strength $\cF$ is given by
\begin{equation} \label{D2 F}
\cF = - B + \dot y_{10} \, \Big( 1 - V^{-1} \cQ \sum_I Z_I^{-1} (\frv_I)^2 \Big)^{-1/2} \bigg( \frac{Q}{V Z_3} \, \dd \psi \wedge \dd z - \frac{\mu V}{Z_3} \, \dd t \wedge \dd z \bigg)~.
\end{equation}
It is more or less straightforward now to check, using the gauge choices
\begin{equation}
B = y_9 \, \dd A^3~, \qquad C_3 = y_5 \, \dd A^1 \wedge \dd y_6 + y_7 \, \dd A^2 \wedge \dd y_8~,
\end{equation}
that all of the (original) equations of motion for $y_5, y_6, y_7, y_8, y_9$ are given by the action \eqref{D2 action}, and that the equation of motion for $y_{10}$ is given by the Bianchi identity for the gauge field $\cF$ \eqref{D2 F}.  The equation of motion for $\cF$ derived from \eqref{D2 action} is automatically satisfied by the expression \eqref{D2 F}.

\subsection{T-duality to type IIB}

Recalling the T-duality rules~\eqref{T-duality rules} and
working in the gauge
\begin{equation}
B = A^3 \wedge \dd y_9~,
\end{equation}
the IIA worldvolume gauge field \eqref{D2 F} is written
\begin{equation}
\cF = - A^3 \wedge \dd y_9 + \dot y_{10} \, \Big( 1 - V^{-1} \cQ \sum_I Z_I^{-1} (\frv_I)^2 \Big)^{-1/2} \bigg( \frac{Q}{V Z_3} \, \dd \psi \wedge \dd z - \frac{\mu V}{Z_3} \, \dd t \wedge \dd z \bigg)~.
\end{equation}
Under T-duality, the gauge field transforms as
\begin{equation}
\tilde \cF = \cF + \dd y_9 \wedge \dd \tilde y_9 = \cF + \dd y_9 \wedge \dd v~,
\end{equation}
and hence the IIB worldvolume gauge field is given by
\begin{equation} \label{D3 F}
\begin{split}
\tilde \cF &= -(\dd v + A^3) \wedge \dd y_9 \\
& \qquad + \dot y_{10} \, \Big( 1 - V^{-1} \cQ \sum_I Z_I^{-1} (\frv_I)^2 \Big)^{-1/2} \bigg( \frac{Q}{V Z_3} \, \dd \psi \wedge \dd z - \frac{\mu V}{Z_3} \, \dd t \wedge \dd z \bigg)~,
\end{split}
\end{equation}
which now has a natural-looking piece that talks to the KKM structure.

There is another complication in that $t$ is now a null coordinate in \eqref{IIB metric A}, due to the $-Z_3^{-1} \, (\dd t + k)$ term in $A^3$.  It turns out that we can just ignore this and calculate.  First, we find the pullback of the metric~\eqref{IIB metric} can be written
\begin{equation} \label{D3 metric A}
\begin{split}
\dd s_{D3} &= -\sqrt{Z_1 Z_2} \Big( (Z_1 Z_2 Z_3)^{-1} - Z_1^{-1} (\frv_1)^2 - Z_2^{-1} (\frv_2)^2 \Big) \dd t^2 - \frac{2 \mu}{Z_3 \sqrt{Z_1 Z_2}} \, \dd t \, \dd \psi \\
& \qquad + \frac{\cQ}{V^2 Z_3 \sqrt{Z_1 Z_2}} \, \dd \psi^2  + \frac{Z_3}{\sqrt{Z_1 Z_2}} \, (\dd v + A^3)^2 + V \sqrt{Z_1 Z_2} \, \dd z^2~,
\end{split}
\end{equation}
and the determinant is
\begin{equation} \label{D3 det A}
- \det g_{D3} = 1 - V^{-1} \cQ Z_1^{-1} (\frv_1)^2 - V^{-1} \cQ Z_2^{-1} (\frv_2)^2~.
\end{equation}
The D3 probe action is, in our case with $\phi = 0$ and $B, C_2, C_0 = 0$:
\begin{equation} \label{D3 action A}
\cS_{D3} = -\tau_{D3} \int_\cW \Big( 1 + \bnorm{\tilde \cF}^2 + \frac14 \bnorm{\tilde \cF \wedge \tilde \cF}^2 \Big)^{1/2} \, \vol_\text{D3} + \tau_{D3} \int_\cW C_4~,
\end{equation}
where now $\cW = \RR_t \times \widetilde\Delta$.  The difference between this and the D2 action is that now \emph{both} $y_9, y_{10}$ are wrapped up into $\tilde \cF$.  So, the $y_9, y_{10}$ equations of motion must come from the equations satisfied by $\tilde \cF$ \eqref{D3 F}.  It is clear that the Bianchi identity
\begin{equation}
\dd \tilde \cF = 0
\end{equation}
still yields the $y_{10}$ equation of motion.  One should expect that the equation of motion for $\tilde \cF$ should yield the $y_9$ equation \eqref{y9 eqn}.  To show this, we can exploit the spectral interchange symmetry to simplify the algebra.

\subsection{Spectral interchange symmetry and gauge field dynamics}

The full IIB metric can be written with manifest spectral-interchange symmetry by appending the torus directions to \eqref{6d metric warner}:
\begin{equation} \label{IIB spec int}
\begin{split}
\dd s_{10}^2 &= - 2 \hat H^{-1} \, (\dd t + \omega) \, \alpha + \hat H^{-3} \Big( \cQ \, \alpha^2 + \gamma^2 \Big) + \hat H \, \dd \vec y \cdot \dd \vec y \\
& \qquad \qquad + \sqrt{\frac{Z_2}{Z_1}} \big( \dd y_5^2 + \dd y_6^2 \big) + \sqrt{\frac{Z_1}{Z_2}} \big( \dd y_7^2 + \dd y_8^2 \big),
\end{split}
\end{equation}
The pullback onto the D3 brane is then simply
\begin{equation} \label{D3 metric 2}
\dd s_{D3}^2 = - 2 \hat H^{-1} \, \dd t \, (i^* \alpha) + \hat H^{-3} \Big( \cQ (i^* \alpha)^2 + (i^* \gamma)^2 \Big) + \hat H \, \dd z^2 + \hat H V^{-1} \sum_{I=1}^2 Z_I^{-1} (\frv_i)^2 \, \dd t^2~.
\end{equation}
The volume form is then easy to compute:
\begin{equation}
\vol_{D3} = \hat H^{-2} \Big( 1 - \cQ V^{-1} \sum_{I=1}^2 Z_I^{-1} (\frv_i)^2 \Big)^{1/2} \, \dd t \wedge i^* \alpha \wedge i^* \gamma \wedge \dd z~.
\end{equation}
The extra factor of $\hat H^{-2}$ is eliminated by expanding out $i^* \alpha, i^* \gamma$:
\begin{equation}
i^* \alpha = V \, \dd v + K^3 \, \dd \psi, \qquad i^* \gamma = (K^3)^2 \tilde \mu \, \dd \psi - V^2 \mu \, \dd v~.
\end{equation}
Using the identity
\begin{equation}
\tilde \mu = - \frac{V}{K^3} \, \mu + \frac{Z_1 Z_2 V}{(K^3)^2}~,
\end{equation}
we then have
\begin{equation}
i^* \alpha \wedge i^* \gamma = \Big( V (K^3)^2 \tilde \mu + V^2 (K^3) \mu \Big) \, \dd v \wedge \dd \psi = \hat H^2 \, \dd v \wedge \dd \psi~,
\end{equation}
and so the volume form is just
\begin{equation}
\vol_{D3} = \Big( 1 - \cQ V^{-1} \sum_{I=1}^2 Z_I^{-1} (\frv_i)^2 \Big)^{1/2} \, \dd t \wedge \dd v \wedge \dd \psi \wedge \dd z~.
\end{equation}
Note that the sum is over the first two $T^2$'s only.

We will need the Hodge dual of $\cF$, for which it is convenient to find orthonormal frames.  Completing the square in \eqref{D3 metric 2} yields
\begin{equation}
\begin{split}
\dd s_{D3}^2 &= - \hat H \cQ^{-1} \Big( 1 - \cQ V^{-1} \sum_{I=1}^2 Z_I^{-1} (\frv_I)^2 \Big) \, \dd t^2 + \hat H^{-3} \cQ \, \big( \alpha - \hat H^2 \cQ^{-1} \, \dd t \big)^2 \\
& \qquad + \hat H^{-3} \, \gamma^2 + \hat H \, \dd z^2~,
\end{split}
\end{equation}
where pullbacks of $\alpha, \gamma$ are now assumed.  This gives us the frames:
\begin{align}
e^0 &= \hat H^{1/2} \cQ^{-1/2} \Big( 1 - \cQ V^{-1} \sum_{I=1}^2 Z_I^{-1} (\frv_I)^2 \Big)^{1/2} \, \dd t~, & e^2 &= H^{-3/2} \, \gamma~, \\
e^1 &= H^{-3/2} \cQ^{1/2} \, \big( \alpha - H^2 \cQ^{-1} \, \dd t \big)~, & e^3 &= H^{1/2} \, \dd z~.
\end{align}
Using these frames, we can write the gauge field strength as
\begin{equation} \label{D3 F frames}
\begin{split}
\cF &= \frac{\dot y_9}{\hat H Z_3} \Big( 1 - \cQ V^{-1} \sum_{I=1}^2 Z_I^{-1} (\frv_I)^2 \Big)^{-1/2} \Big( \cQ \, e^0 \wedge e^1 - \mu V \cQ^{1/2} \, e^0 \wedge e^2 \Big) \\
& \qquad + \frac{\dot y_{10}}{\hat H Z_3} \Big( 1 - \cQ V^{-1} \sum_{I=1}^3 Z_I^{-1} (\frv_I)^2 \Big)^{-1/2} \Big( - \cQ \, e^2 \wedge e^3 - \mu V Q^{1/2} \, e^1 \wedge e^3 \Big)~.
\end{split}
\end{equation}
Note that the two prefactors $(\cdots)^{-1/2}$ are different, due to the range of the sums.  The Hodge dual of $\cF$ is given by
\begin{equation} \label{D3 star F frames}
\begin{split}
\hodge_{D3} \cF &= - \frac{\dot y_{10}}{\hat H Z_3} \Big( 1 - \cQ V^{-1} \sum_{I=1}^3 Z_I^{-1} (\frv_I)^2 \Big)^{-1/2} \Big( \cQ \, e^0 \wedge e^1 - \mu V \cQ^{1/2} \, e^0 \wedge e^2 \Big) \\
& \qquad + \frac{\dot y_9}{\hat H Z_3} \Big( 1 - \cQ V^{-1} \sum_{I=1}^2 Z_I^{-1} (\frv_I)^2 \Big)^{-1/2} \Big( - \cQ \, e^2 \wedge e^3 - \mu V Q^{1/2} \, e^1 \wedge e^3 \Big)~.
\end{split}
\end{equation}
Note that the $\dot y_9, \dot y_{10}$ \emph{and} the prefactors $(\cdots)^{-1/2}$ have switched places.  We note that the na\"ive dual equation (which is \emph{not} an equation of motion for \eqref{D3 action A}),
\begin{equation}
\dd \hodge_{D3} \cF = 0~,
\end{equation}
fails to give the original $y_9$ equation \eqref{y9 eqn}, precisely because these prefactors are switched.  We will show that the correct equations of motion, derived from \eqref{D3 action A}, switch these factors back via a bit of seemingly-magical algebra, and thus give the correct equation for $y_9$ (as should be expected by T-duality).

\subsubsection{Varying the action}

There are two ways we can write the DBI action on the D3 branes.  The more traditional way is
\begin{equation}
\cS_{D3} = - \tau_{D3} \int_\cW \dd^4 \xi \, \sqrt{- \det M_{\mu\nu}}
~, \qquad M_{\mu\nu} \equiv (i^*g)_{\mu\nu} + \cF_{\mu\nu}~,
\end{equation}
where the matrix $M_{\mu \nu}$ is the usual combination of symmetric and antisymmetric parts.  Varying with respect to the gauge field, one straightforwardly obtains the equation of motion
\begin{equation}
\partial_\mu \big( \sqrt{- \det M} \, M^{[\mu\nu]} \big) = 0~,
\end{equation}
where $M^{[\mu\nu]}$ is the antisymmetric part of the inverse of $M_{\mu\nu}$.  The catch here is that it is not so straightforward to invert $M_{\mu\nu}$.  One way to get an explicit equation of motion for $\cF_{\mu\nu}$ is to apply the Cayley-Hamilton theorem which allows one to write the inverse of a matrix in terms of its characteristic polynomial.

Another (and equivalent) route uses the version of the action~\eqref{alt DBI}:
\begin{equation} 
\cS_{D3} = -\tau_{D3} \int_\cW \Big( 1 + \bnorm{\cF}^2 + \frac14 \bnorm{\cF \wedge \cF}^2 \Big)^{1/2} \, \vol_\text{D3}~,
\end{equation}
for which it is slightly more work to compute the variation, but one ends up with an explicit equation of motion
\begin{equation}
\dd \cG = 0~,
\end{equation}
in terms of the ``DBI-dual" field strength
\begin{equation}
\cG \equiv \Big( 1 + \bnorm{\cF}^2 + \frac14 \bnorm{\cF \wedge \cF}^2 \Big)^{-1/2} \bigg[ \hodge_{D3} \cF + \frac12 \Big( \hodge_{D3} (\cF \wedge \cF) \Big) \, \cF \bigg]~.
\end{equation}
We emphasize that the second term contains a function multiplying $\cF$, and not $\hodge_{D3} \cF$.

Next we compute everything that appears in the definition of $\cG$.  First we need
\begin{equation}
\begin{split}
\cF \wedge \cF &=  - 2 \frac{\cQ}{V Z_3} \dot y_9 \dot y_{10} \Big( 1 - \cQ V^{-1} \sum_{I=1}^2 Z_I^{-1} (\frv_I)^2 \Big)^{-1/2} \times \\
& \qquad \qquad \qquad \times \Big( 1 - \cQ V^{-1} \sum_{I=1}^3 Z_I^{-1} (\frv_I)^2 \Big)^{-1/2} \, e^0 \wedge e^1 \wedge e^2 \wedge e^3~.
\end{split}
\end{equation}
It is then straightforward to show that
\begin{equation}
1 + \bnorm{\cF}^2 + \frac14 \bnorm{\cF \wedge \cF}^2 = \frac{ \Big( 1 - \cQ V^{-1} \sum_{I=1}^2 Z_I^{-1} (\frv_I)^2 - \cQ V^{-1} Z_3^{-1} (\dot y_9^2) \Big)^2 }{ \Big( 1 - \cQ V^{-1} \sum_{I=1}^2 Z_I^{-1} (\frv_I)^2 \Big) \Big( 1 - \cQ V^{-1} \sum_{I=1}^3 Z_I^{-1} (\frv_I)^2 \Big) }~.
\end{equation}
Again note the range of the sums.  Finally, after a bit more algebra, one can show that
\begin{equation} \label{D3 G frames}
\begin{split}
\cG &= -\frac{\dot y_{10}}{\hat H Z_3} \Big( 1 - \cQ V^{-1} \sum_{I=1}^2 Z_I^{-1} (\frv_I)^2 \Big)^{-1/2} \Big( \cQ \, e^0 \wedge e^1 - \mu V \cQ^{1/2} \, e^0 \wedge e^2 \Big) \\
& \qquad + \frac{\dot y_9}{\hat H Z_3} \Big( 1 - \cQ V^{-1} \sum_{I=1}^3 Z_I^{-1} (\frv_I)^2 \Big)^{-1/2} \Big( - \cQ \, e^2 \wedge e^3 - \mu V Q^{1/2} \, e^1 \wedge e^3 \Big)~,
\end{split}
\end{equation}
where, as promised, the $(\cdots)^{-1/2}$ factors have switched places relative to \eqref{D3 star F frames}.

Plugging back in our frames, and the definitions of $\alpha, \gamma$, we find a remarkable symmetry:
\begin{align}
\begin{split} \label{D3 F final}
\cF &= - \dot y_9 \, (\dd v + A^3) \wedge \dd t \\
& \qquad + \dot y_{10} \, \Big( 1 - \cQ V^{-1} \sum_{I=1}^3 Z_I^{-1} (\frv_I)^2 \Big)^{-1/2} \bigg( \frac{Q}{V Z_3} \, \dd \psi \wedge \dd z - \frac{\mu V}{Z_3} \, \dd t \wedge \dd z \bigg)~,
\end{split} \\
\begin{split} \label{D3 G final}
\cG &= \dot y_{10} \, (\dd v + A^3) \wedge \dd t \\
& \qquad + \dot y_9 \, \Big( 1 - \cQ V^{-1} \sum_{I=1}^3 Z_I^{-1} (\frv_I)^2 \Big)^{-1/2} \bigg( \frac{Q}{V Z_3} \, \dd \psi \wedge \dd z - \frac{\mu V}{Z_3} \, \dd t \wedge \dd z \bigg)~.
\end{split}
\end{align}
Thus we see that $\dd \cG = 0$ gives the $y_9$ equation \eqref{y9 eqn} in a similar way that $\dd \cF = 0$ gives the $y_{10}$ equation \eqref{y10 eqn}.


\vskip 2cm
\bibliographystyle{JHEP}
\bibliography{Fuzz}

\end{document}